\documentclass[useAMS,usenatbib]{mn2e}

\usepackage{amsmath}
\usepackage{natbib}
\usepackage{graphicx}
\usepackage{amssymb}
\usepackage{amsmath}
\usepackage{mathrsfs}
\usepackage[usenames]{color}
\title {Iron Line Profiles in Strong Gravity}
\author [Kris Beckwith and Chris Done]{Kris Beckwith$^1$ and Chris Done$^1$ \\
$^1$Department of Physics, University of Durham, South Road, Durham DH1 3LE, UK}
\date{Released 2004 Xxxxx XX}

\pagerange{\pageref{firstpage}--\pageref{lastpage}} \pubyear{2004}

\def\LaTeX{L\kern-.36em\raise.3ex\hbox{a}\kern-.15em
    T\kern-.1667em\lower.7ex\hbox{E}\kern-.125emX}

\begin{document}

\label{firstpage}

\maketitle

\begin{abstract}

We describe a new code which can accurately calculate the relativistic
effects which distort the emission from an accretion disc around a
black hole. We compare our results for a disk which extends from the
innermost stable orbit to $20r_{g}$ in both Schwarzschild and maximal
($a=0.998$) Kerr spacetimes with the two line profile codes which are
on general release in the XSPEC spectral fitting package. These models
generally give a very good description of the relativistic smearing of
the line for this range of radii. However, these models have some
limitations. In particular we show that the assumed form of the {\em
angular} emissivity law (limb darkening or brightening) can make
significant changes to the derived line profile where lightbending is
important. This is {\em always} the case for extreme Kerr spacetimes
or high inclination systems, where the observed line is produced
from a very large range of different emitted angles. In these
situations the assumed angular emissivity can affect the derived {\em
radial} emissivity. The line profile is not simply determined by the
well defined (but numerically difficult) physical effects of strong
gravity, but is also dependent on the poorly known astrophysics of the
disc emission.

\end{abstract}


\section{Introduction}

Material in an accretion disk around a black hole is orbiting at
high velocity, close to the speed of light, in a strong gravitational
potential. Hence its emission is distorted by doppler shifts, length
contraction, time dilation, gravitational redshift and
lightbending. The combined impact of these special and general
relativistic effects was first calculated in the now seminal paper of
\cite{C75}, where he used  a \emph{transfer function} to describe the
relativistic effects. The observed spectrum from an accretion disc
around a Kerr black hole is
the  convolution of  this with the intrinsic disc continuum emission.

While such models have been used to try to determine the gravitational
potential from the observed accretion disk spectra
(e.g. \citealt{LN89,EMH91,E93,M00,GME01}), 
these attempts suffer from our limited knowledge of the
spectral shape of the intrinsic accretion disk emission (see e.g. the
review by \citealt{B02}). It is much easier to determine the relativistic
effects from a {\em sharp} spectral feature, such as the iron
fluorescence line expected from X-ray illumination of an accretion disc
\citep{F89}. An originally narrow atomic transition is
transformed into broad, skewed profile whose shape is given {\em
directly} by the transfer function.

Observationally, evidence for a relativistically smeared iron line
first came from the ASCA observation of the active galactic nuclei
(AGN) MCG-6-30-15 \citep{T95}. Further observations showed
evidence for the line profile being so broad as to require a maximally
spinning black hole \citep{I96}. More recent data from
XMM are interpreted as showing that the line is even wider than 
expected from an extreme Kerr disk, requiring direct 
extraction of the spin energy from the central black hole 
as well as the immense gravitational potential \citep{W01}.

Such results are incredibly exciting, but X-ray spectral fitting is
not entirely unambiguous. There is a complex reflected continuum as
well as the line (\citealt{NKK00,BRF01}). For an ionised disk (as
inferred for MCG-6-30-15) the current models in general use ({\tt
pexriv} in the {\sc XSPEC} spectral fitting package) are probably
highly incomplete \citep{RFY99}. Complex ionised absorption also
affects AGN spectra (e.g. \citealt{K02}) and the illuminating
continuum itself can have complex curvature rather than being a simple
power law.

However, in MCG-6-30-15 these issues have been examined in detail, and
the results on the dramatic line width appear robust
(\citealt{VF03,R04}).  Thus there is a clear requirement that the
extreme relativistic effects are well modelled. There are two models
which are currently widely available to the observational community,
within the \texttt{XSPEC} spectral fitting package, \texttt{diskline}
(based on \citealt{F89}) and \texttt{laor} \citep{L91}. The analytic
\texttt{diskline} code models the line profile from an accretion disc
around a Schwarzschild black hole (so of course cannot be used to
describe the effects in a Kerr geometry). Also, it does not include
the effects of lightbending (although \citealt{F89} outline a scheme
for incorporating this) and hence does not accurately calculate all
the relativistic effects for $r< 20r_{g}$ (where $r_g=GM/c^2$).  By
contrast, the \texttt{laor} model numerically calculates the line
profile including lightbending for an extreme Kerr black hole, but
uses a rather small set of tabulated transfer functions which limit
its resolution and accuracy (see Section 3.3).

While there are other relativistic codes in the literature which do
not suffer from these limitations, these are not generally readily
and/or easily available for observers to use. There is a clear need
for a fast, accurate, high resolution code which can be used to fit
data from the next generation of satellites. In this paper we describe
our new code for computing the relativistic iron line profile in both
the Schwarzschild and Kerr metrics. We compare this with the
\texttt{diskline} and \texttt{laor} models in \texttt{XSPEC} for discs
which extend down to the last stable orbit in their respective
spacetimes, and highlight both the strengths and
limitations of these previous models.

\section{Calculating Strong Gravitational Effects}

\begin{figure}
  \leavevmode
  \begin{center}
  \includegraphics[width=0.9\columnwidth]{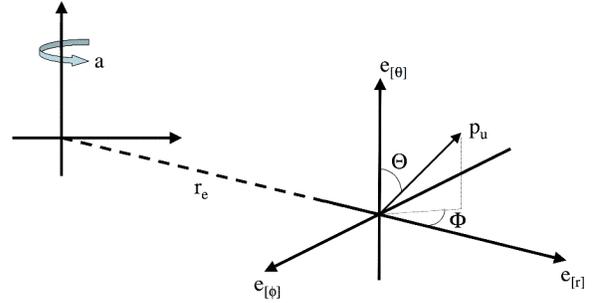}
  \caption{The coordinate system used for the disc. The emission
is defined in the rest frame of the disc material. The polar and azimuthal 
emission angles $\Theta, \Phi$ are obtained by taking the dot-products of the 
photon four-momentum with the basis vectors of this frame, where $\mu_{e}= 
\cos\Theta$.  This disc frame can be connected to the frame which co-rotates with 
the black hole spacetime via a simple boost which depends on the velocity 
structure of the disc.}
  \label{fig:2.1.1}
  \end{center}
\end{figure}

\begin{figure}
  \leavevmode
  \begin{center}
  \includegraphics[width=0.9\columnwidth]{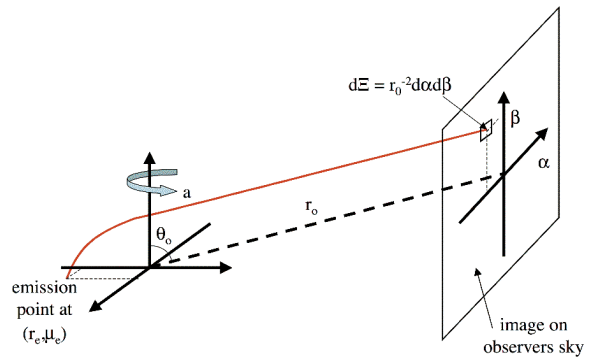}
  \caption{Diagram showing the link between the observers frame of reference and 
the global coodinate system defined by the black hole. Photons that are emitted 
from the disc at some  distance $r_e$ from the hole are seen at coordinates 
$\alpha,\beta$ on the image of the disc at the observer.}
  \label{fig:2.1.2}
  \end{center}
\end{figure}

We follow the standard approach (e.g. \citealt{C75,F89,F97}) and calculate
an infinitesimal amount of flux, $dF_{o}$ observed at energy, $E_{o}$
due to a patch on the disk which subtends a solid angle $d\Xi$ 
on the image of the disc at the observer (see Fig. \ref{fig:2.1.1} and 
\ref{fig:2.1.2}).
\begin{align}
  \label{eqn:2.1.1}
    dF_{o} \left( E_{o} \right) = I_{o} \left( E_{o} \right) d\Xi = g^{3} I_{e} 
\left( E_{e} \right) d\Xi
\end{align}
where the redshift factor $g=E_{o}/E_{e}$ and specific intensity in the
observers and emitters frame, $I_{o}$ and $I_{e}$ are related
through the relativistic invariant $I/\nu^3$. For an emission line with rest 
energy
$E_{int}$, then $I_{e} \left( E_{e} \right) = \varepsilon \left( r_{e},
\mu_{e} \right)\delta \left( E_{e} - E_{int} \right)$, where
$\varepsilon \left( r_{e},\mu_{e} \right)$ is the emissivity, which
can be a function of the radius, $r_{e}$ and angle, $\mu_{e}$ at which the photon 
is emitted (as defined in Figure \ref{fig:2.1.1}). The infinitesimal flux becomes
\begin{align}
  \label{eqn:2.1.2}
    dF_{o} \left( E_{o} \right) =  g^{4} \varepsilon \left(
    r_{e}, \mu_{e} \right)
    \delta \left( E_{o} - gE_{int} \right) d\Xi
\end{align}
The total flux can be obtained by integrating over all the entire image of
the disk in the observers sky. We can write $d\Xi=d\alpha d\beta/r_{o}^2$ 
where $\alpha,\beta$ are the $x,y$ coordinates of the image of the
disc at the observer with coordinates $\left( r_{o}, \theta_{o} \right)$ (see 
Figure \ref{fig:2.1.2}), such that 
\begin{align}
  \label{eqn:2.1.3}
    F_{o} \left( E_{o} \right) =  \frac{1}{r_{o}^{2}} \int \int  g^{4}
\varepsilon \left( r_{e},\mu_{e} \right) \delta \left( E_{o} - gE_{int} \right) 
d\alpha d\beta
\end{align}
The $\alpha,\beta$ position of the image of the disc section is 
related to the conserved quantities, $\lambda,q$ which describe the contributions 
to the photons angular momentum from the radial, polar and azimuthal directions 
(\citealt{FP99}), via:
\begin{align}
  \label{eqn:2.1.4}
    \alpha=- \frac{\lambda}{\sin \theta_{o}} \\
    \beta = \pm \sqrt {q^{2} - \lambda^{2} \cot^{2} \theta_{o}}
\end{align}
For a thin, Keplerian disc, these constants of motion can be written in
terms of the redshift factor of the photon, $g$ and the radius of emission, 
$r_{e}$ and angle of emission, $\mu_{e}$ of the photon (as previously defined): 
\begin{align}
  \label{eqn:2.1.5}
    \lambda = \frac{1}{\Omega} \left( 1 - \frac{e^{-\psi}}{g} \right) \\
    q  =  \frac{r_{e} \mu_{e}}{g}
\end{align}
Here, $\Omega$ describes the azimuthal velocity profile of the emitting region and 
$e^{-\psi}$ is the 'redshift function' (\citealt{F97,MPHD}), which for a 
geometrically 
thin, Keplerian disc located in the equatorial plane are given by:
\begin{align}
  \label{eqn:2.1.6}
     \Omega = \frac{1}{a + \sqrt {r_{e}^{3}}} \\
     e^{\psi} = \left[ 1 - \frac{2}{r_{e}} \left( 1 - a\Omega \right)^{2}
     - \left( r_e^{2} + a^{2} \right) \Omega^{2} \right]^{-\frac{1}{2}}
 \end{align}
Thus the problem reduces to finding the area on the observers sky subtended
by all parts of the disc which contribute to a given $E_{o}$.

The photons (null geodesics) that link the accretion disc with the 
observer can only be found by determining the full general relativistic light 
travel paths which link the disc to the
observer. These null geodesics are given by solutions of the geodesic equations 
(\citealt{C68,MTW73,C83}), which  can be obtained numerically (e.g. \citealt{KVP92}), but 
can also be given in terms of analytic functions (\citealt{RB94,APHD,C98}), which 
enable them to be solved quickly and with arbitrary accuracy.

\begin{figure}
  \leavevmode
  \begin{center}
  \includegraphics[width=0.9\columnwidth]{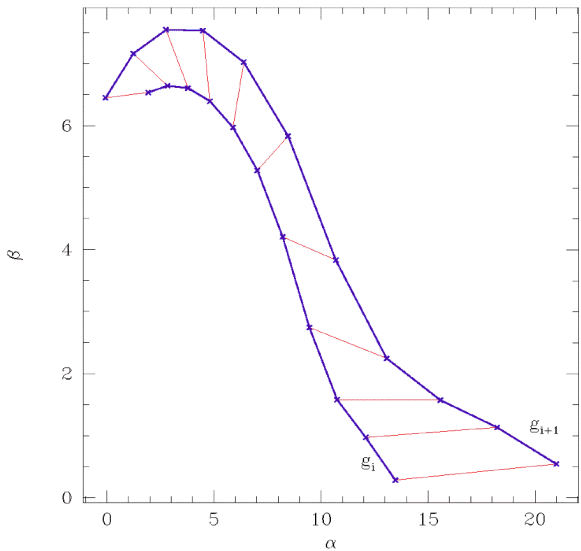}
  \caption{Heavy blue lines denote adjacent contours of constant redshift, $g_{i} 
< g_{i+1}$ on the 
observers sky (the $\alpha, \beta$-plane) that define the area of the redshift bin  
$g = 
g_{i} + \frac{1}{2}  \left( g_{i+1} - g_{i} \right)$ with width $dg = g_{i+1} - 
g_{i}$. Red narrow lines show examples of the divisions used to create a set of 
meshed trapezoids that enable the area of the redshift bin to be determined. For 
the purposes of clarity, this mesh is far coarser than that used in the 
calculations.}
  \label{fig:2.1.3}
  \end{center}
\end{figure}

We develop a technique based solely on the image of the accretion disc
in the $\alpha, \beta$ plane, which defines the flux received by the observer, 
similar to that employed by \cite{CCF03}. This allows us to generate high 
resolution, accurate line profiles numerically while avoiding the issues 
surrounding the partial derivatives of the geodesic equations (\citealt{V93}). We 
use the analytic solutions of the geodesic equations as tabulated by \cite{RB94} 
to find the complete set of light travel paths that link the accretion disk and 
the observer at $\left( r_{o}, \theta_{o} \right)$. We sort these by redshift 
factor, and use adaptive griding to find the boundarys on the $\left( \alpha,\beta 
\right)$-plane for all lines of constant $g$.

Two adjacent boundaries, $g_{i}$ and $g_{i+1}$, therefore define the area of the 
redshift bin $g = g_{i} + \frac{1}{2}  \left( g_{i+1} - g_{i} \right)$ with width 
$dg = g_{i+1} - g_{i}$ when projected onto the $\left( \alpha, \beta \right)$-
plane (as is shown 
in Figure \ref{fig:2.1.3}). We can simply determine the area of this region by 
dividing it up into a set of tessellating trapezoids, as shown in Figure 
{\ref{fig:2.1.3}, the area of 
of each of which can be determined by a simple geometric formula. The final area 
of the redshift 
bin is determined by summing together the contributions from all such trapezoids 
internal 
to $\left( g_{i}, g_{i+1} \right)$. Each individual trapezoid is small, so that 
there is no significant
change in $r_{e}$ or $\mu_{e}$ (though this is not necessarily true
across the total area $d\alpha d\beta$). The emissivity law can be 
convolved into the calculation using the emission coordinates at the centre of 
each trapezoid to weight its area before performing the summation over all 
trapeziods. This approach allows us to calculate line profiles at high spectral 
resolution on timescales of a few minutes on a 2GHz desktop PC.

We have extensively tested the routines that calculate the null geodesic paths 
against those supplied by Eric Agol (\citealt{APHD}) and have found them to be 
indistinguishable. We have also compared the line profiles generated by our code 
to those presented previously in the literature, in particular those generated 
from the code described by \cite{F97} and have again found them to be 
indistinguishable.

\section{Highly Relativistic Line Profiles}

\subsection{Introduction}

\begin{figure*}
  \leavevmode
  \begin{center}
  \begin{tabular}{cc}
  \includegraphics[width=0.45\textwidth]{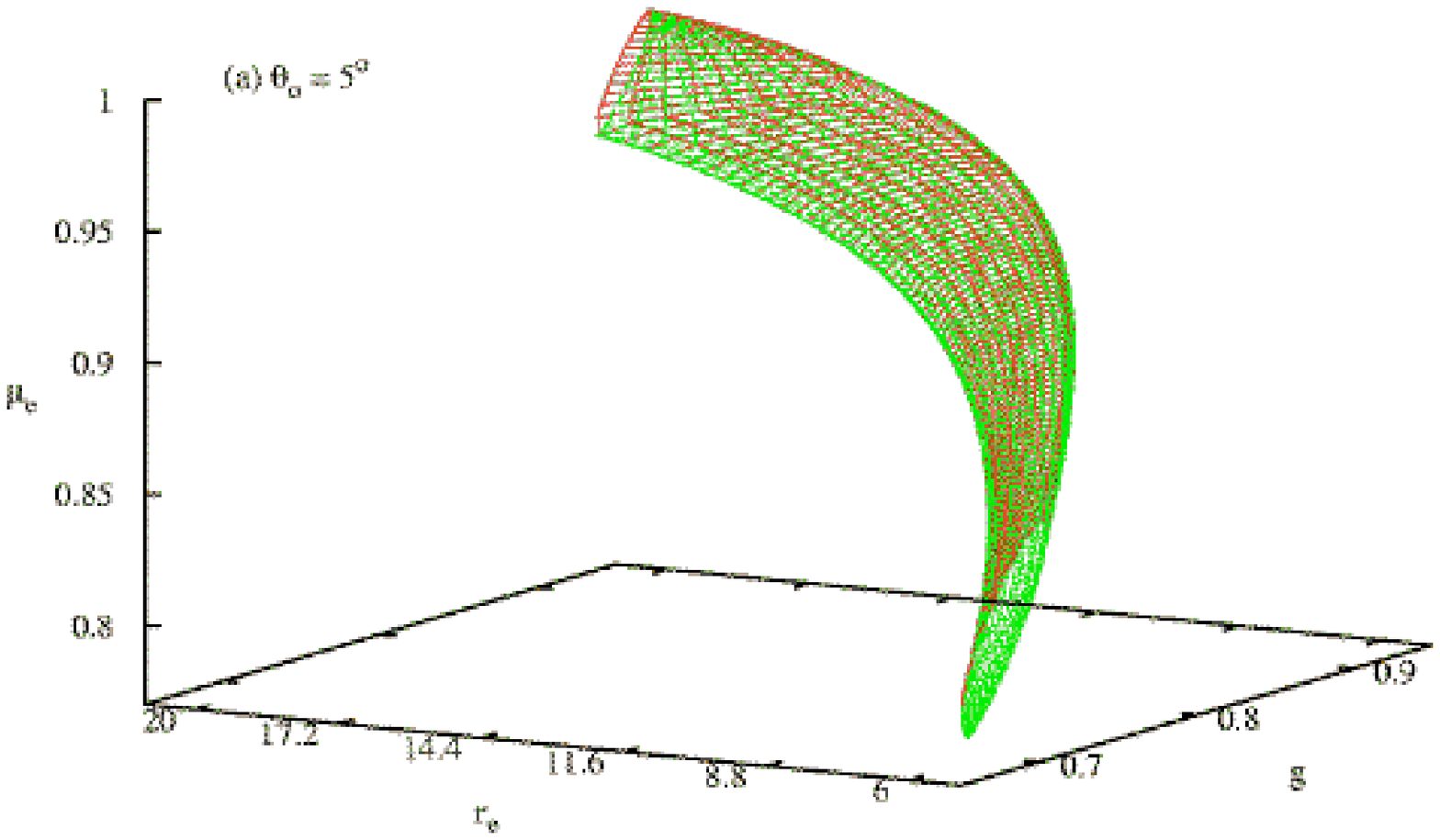}
  &
  \includegraphics[width=0.45\textwidth]{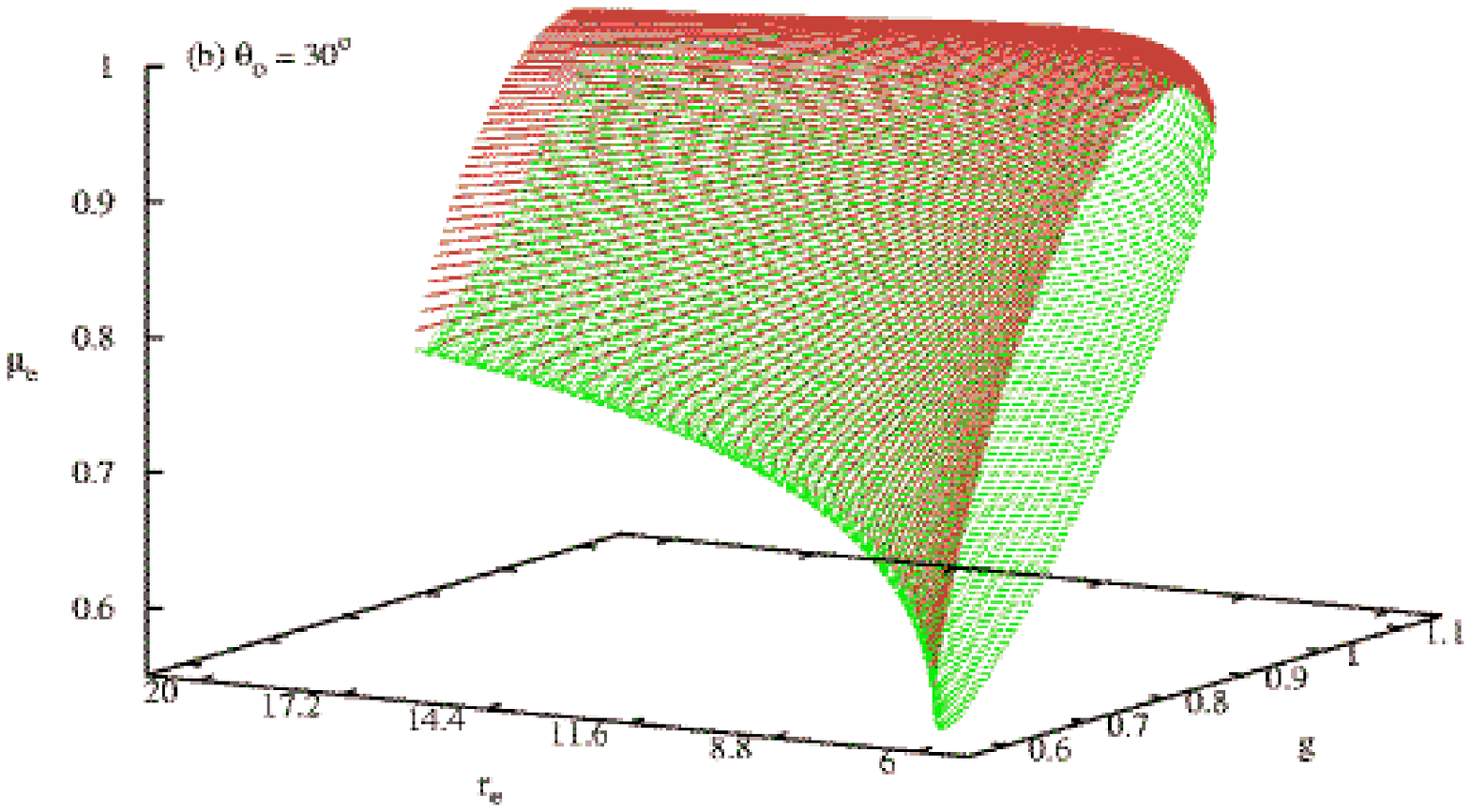}
  \\
  \includegraphics[width=0.45\textwidth]{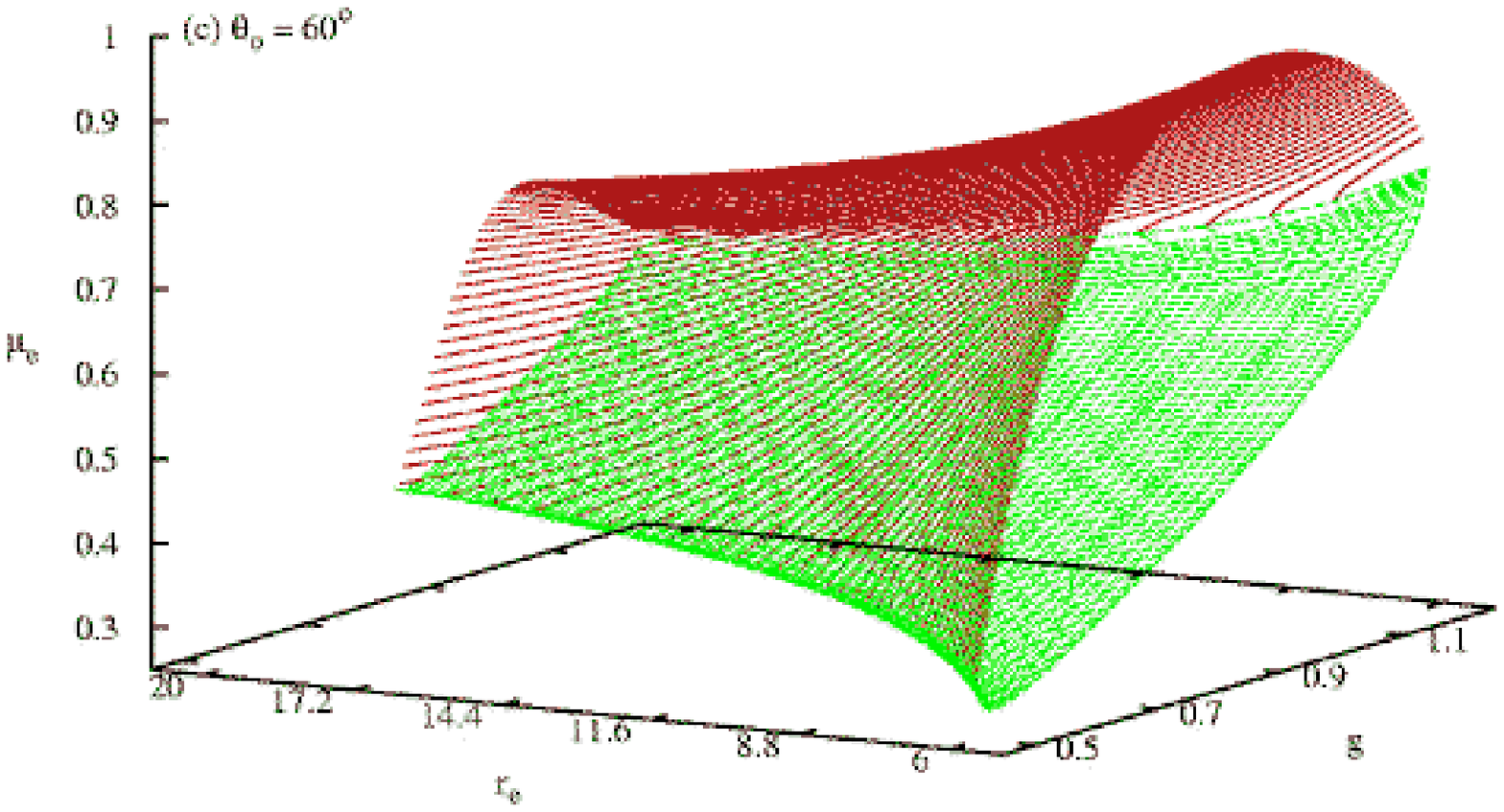}
  &
  \includegraphics[width=0.45\textwidth]{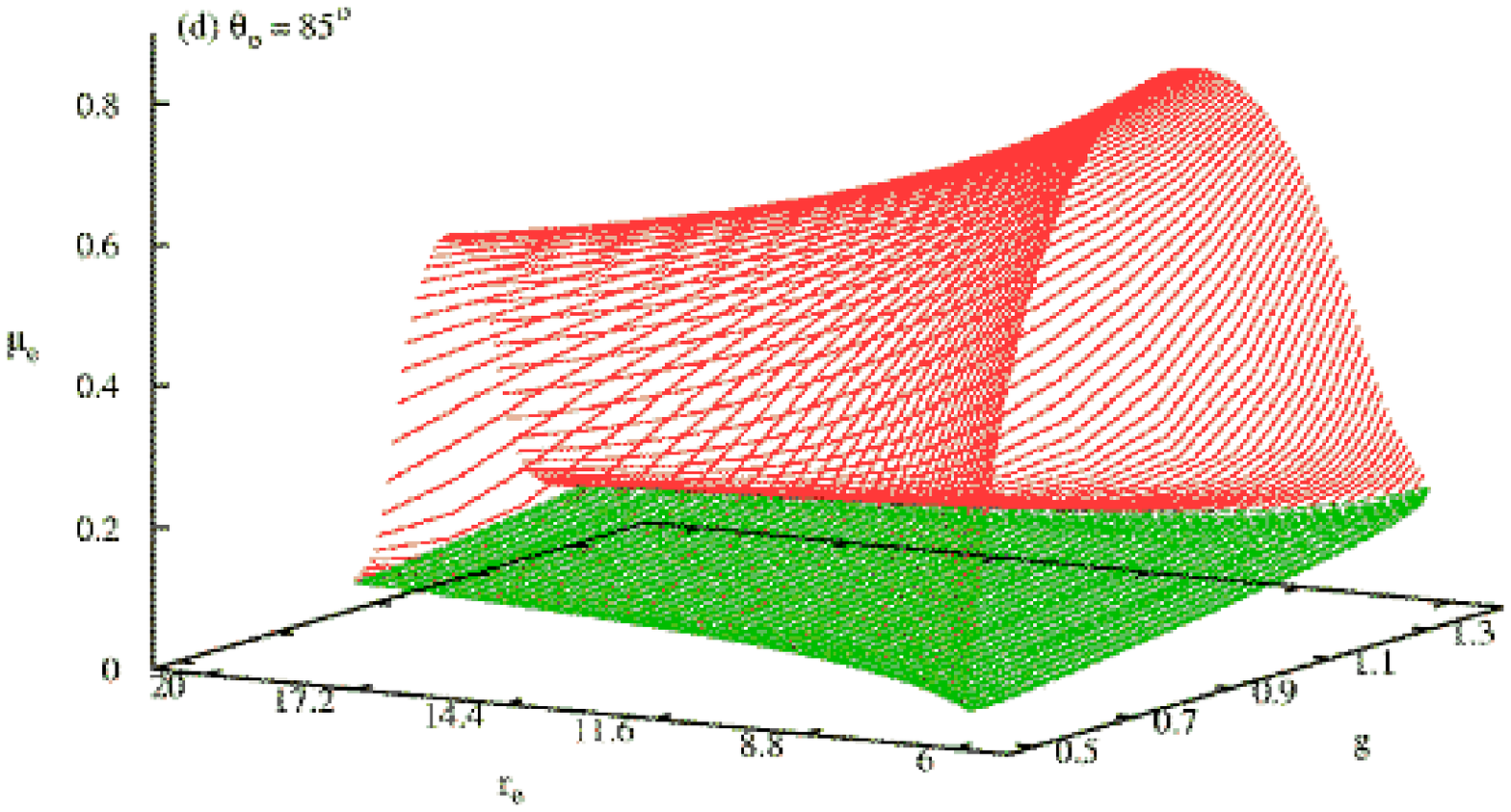}
  \end{tabular}
  \caption{Surfaces in the $\left( \mu_{e}, r_{e}, g \right)$ parameter space 
describing 
geodesics that reach an observer at a given inclination for a standard accretion 
disk 
around a Schwarzschild black hole. Notice that, for every $\left( r_{e}, g 
\right)$ pair, there are two geodesics that reach any given observer, 
corresponding respectively 
to geodesics that are emitted from the side of the disc closest to the observer 
(lower surface) and geodesics that are emitted from the opposite side of the disc 
to the observer (upper surface). These are the two geodesics referred to by 
Cunningham (1975).}
  \label{fig:3.1.1}
  \end{center}
\end{figure*}
  
\begin{figure*}
  \leavevmode
  \begin{center}
  \begin{tabular}{cc}
  \includegraphics[width=0.45\textwidth]{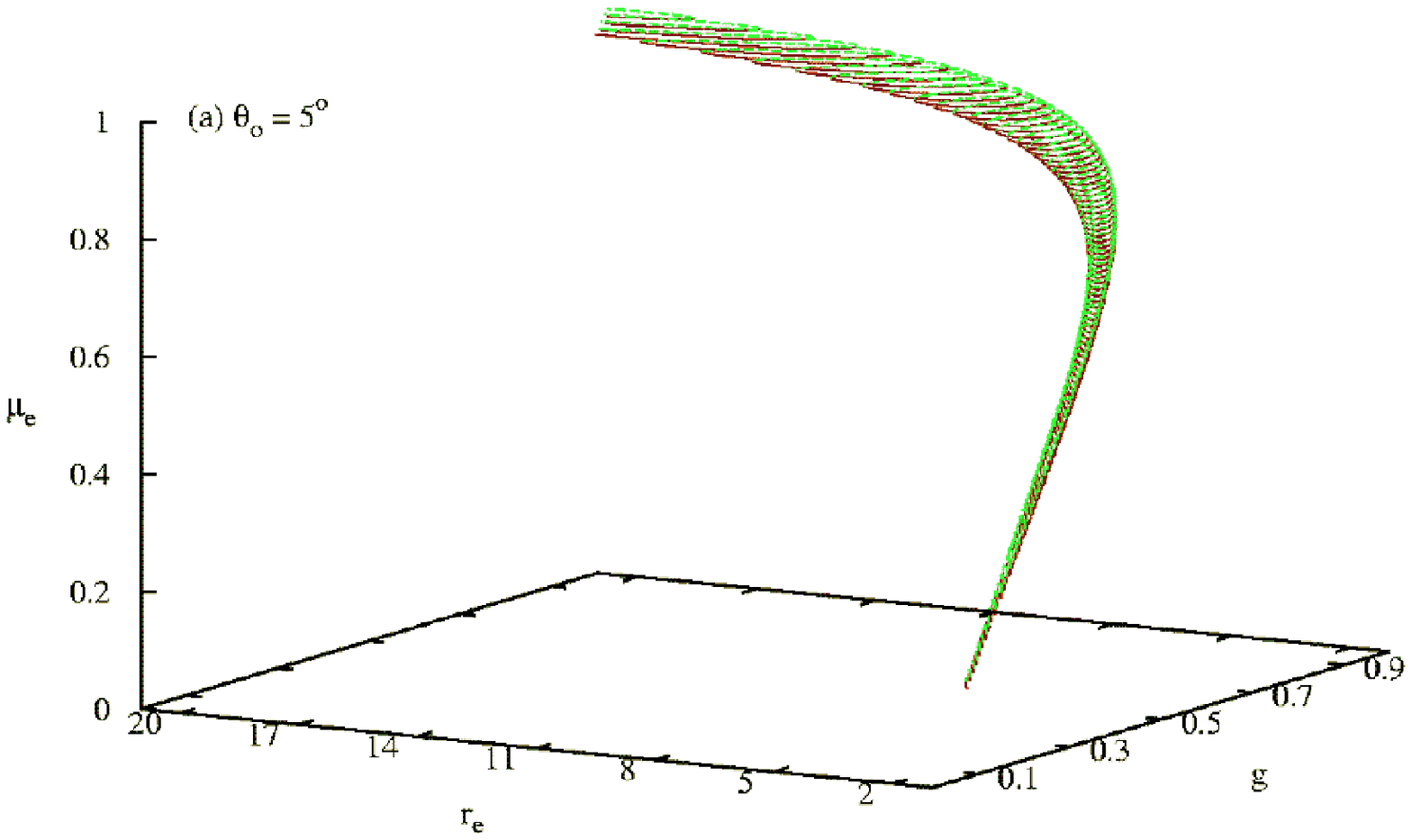}
  &
  \includegraphics[width=0.45\textwidth]{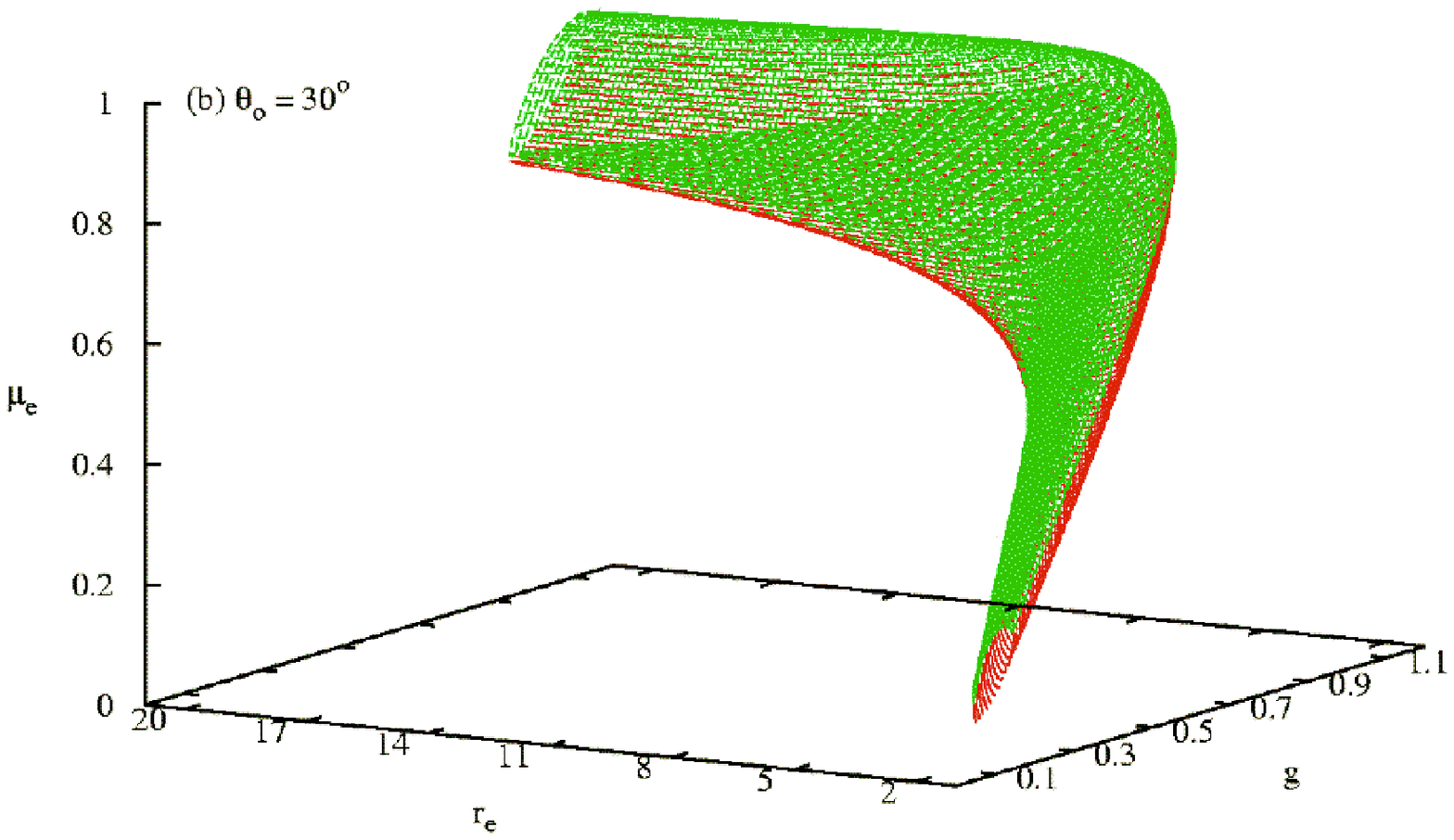}
  \\
  \includegraphics[width=0.45\textwidth]{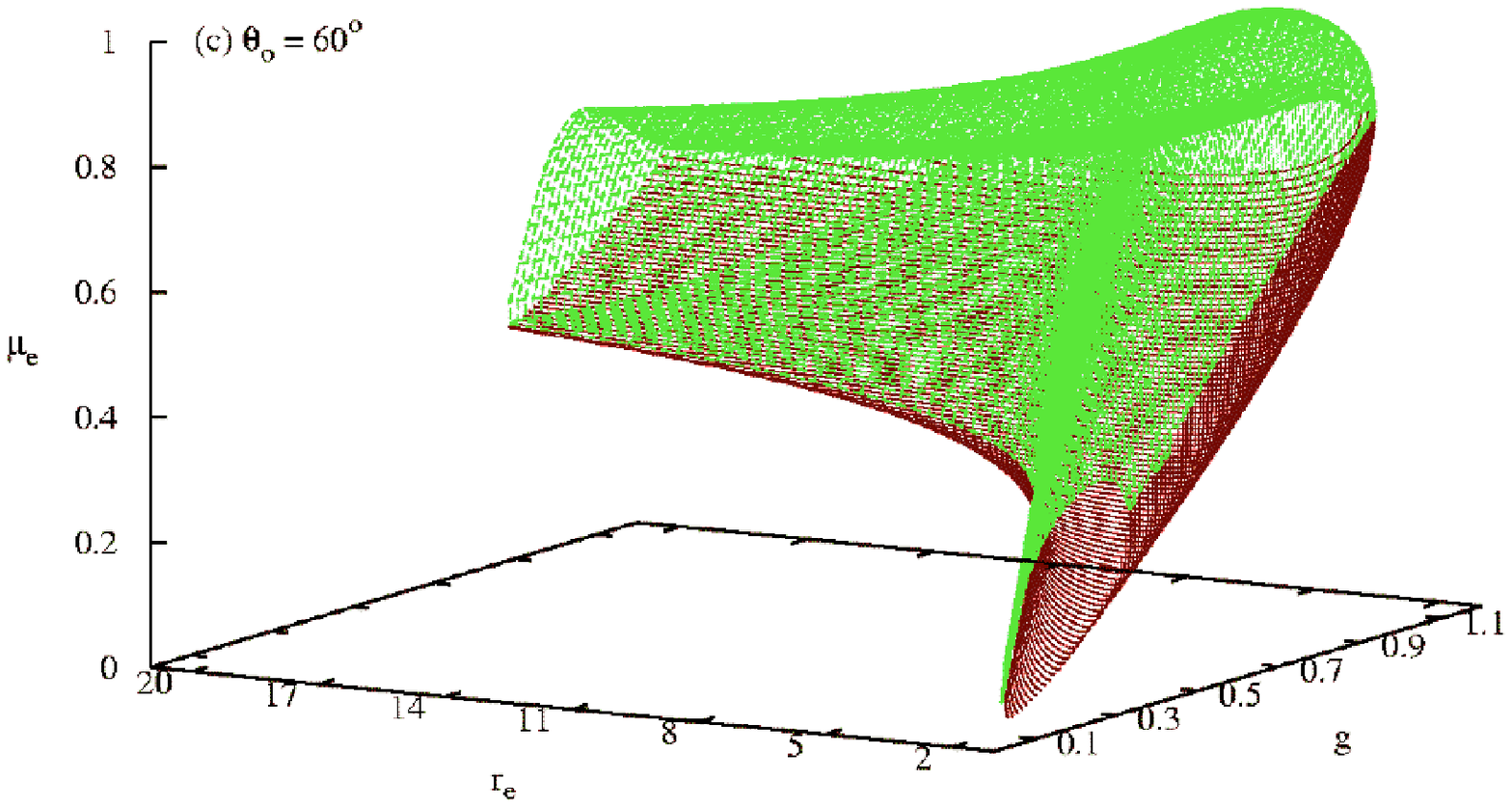}
  &
  \includegraphics[width=0.45\textwidth]{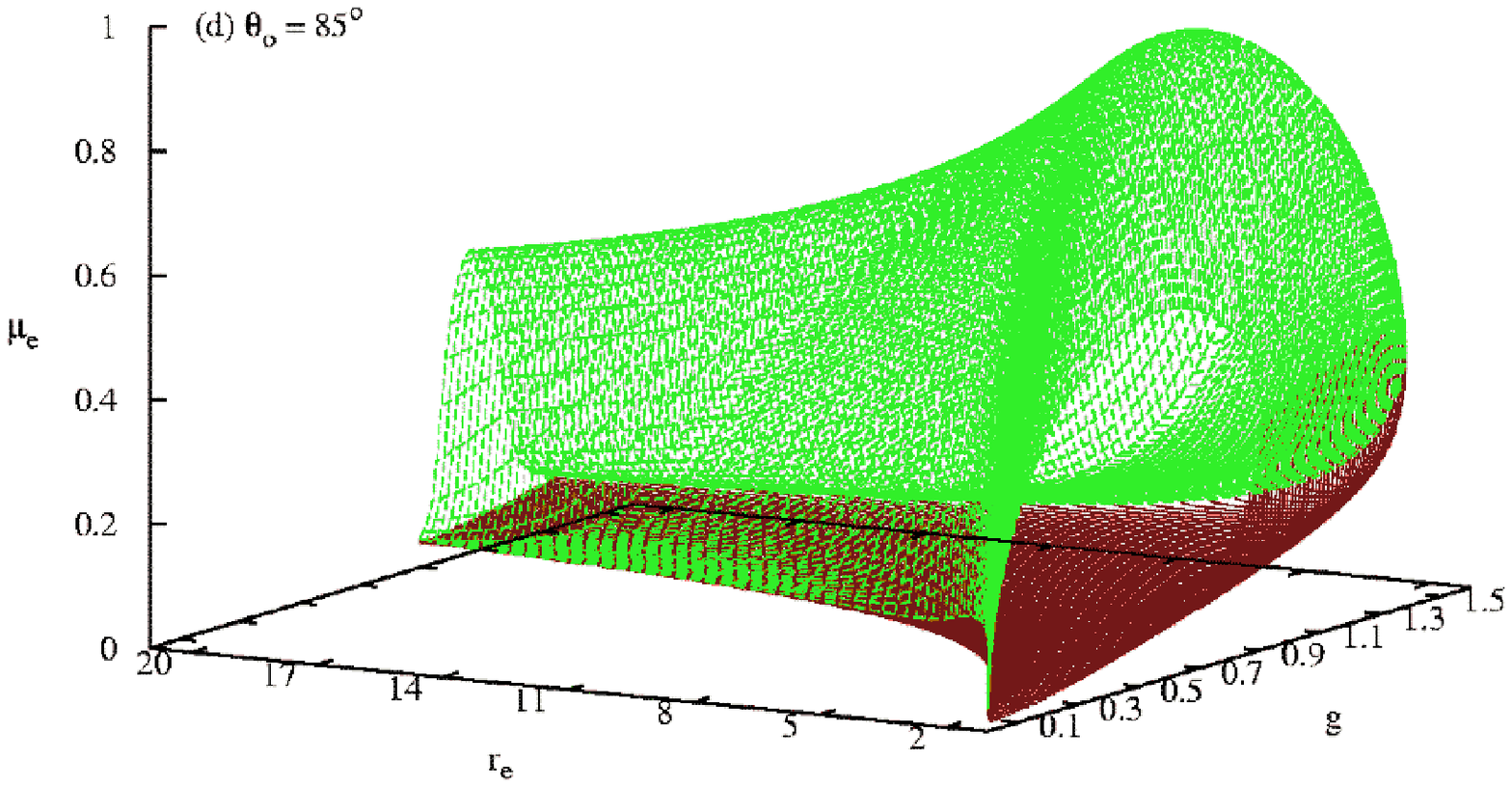}
  \end{tabular}
  \caption{As in Figure \ref{fig:3.1.1} for a maximal ($a=0.998$) Kerr black hole. 
These surfaces are exhibit a far greater complexity than those in the 
Schwarzschild case. The range of accessible redshift is increased for a given 
disc-observer system, whilst the range of emission angles is from zero to unity 
for all inclinations.}
  \label{fig:3.1.2}
  \end{center}
\end{figure*}

\begin{figure*}
  \leavevmode
  \begin{center}
  \begin{tabular}{cc}
  \includegraphics[width=0.45\textwidth]{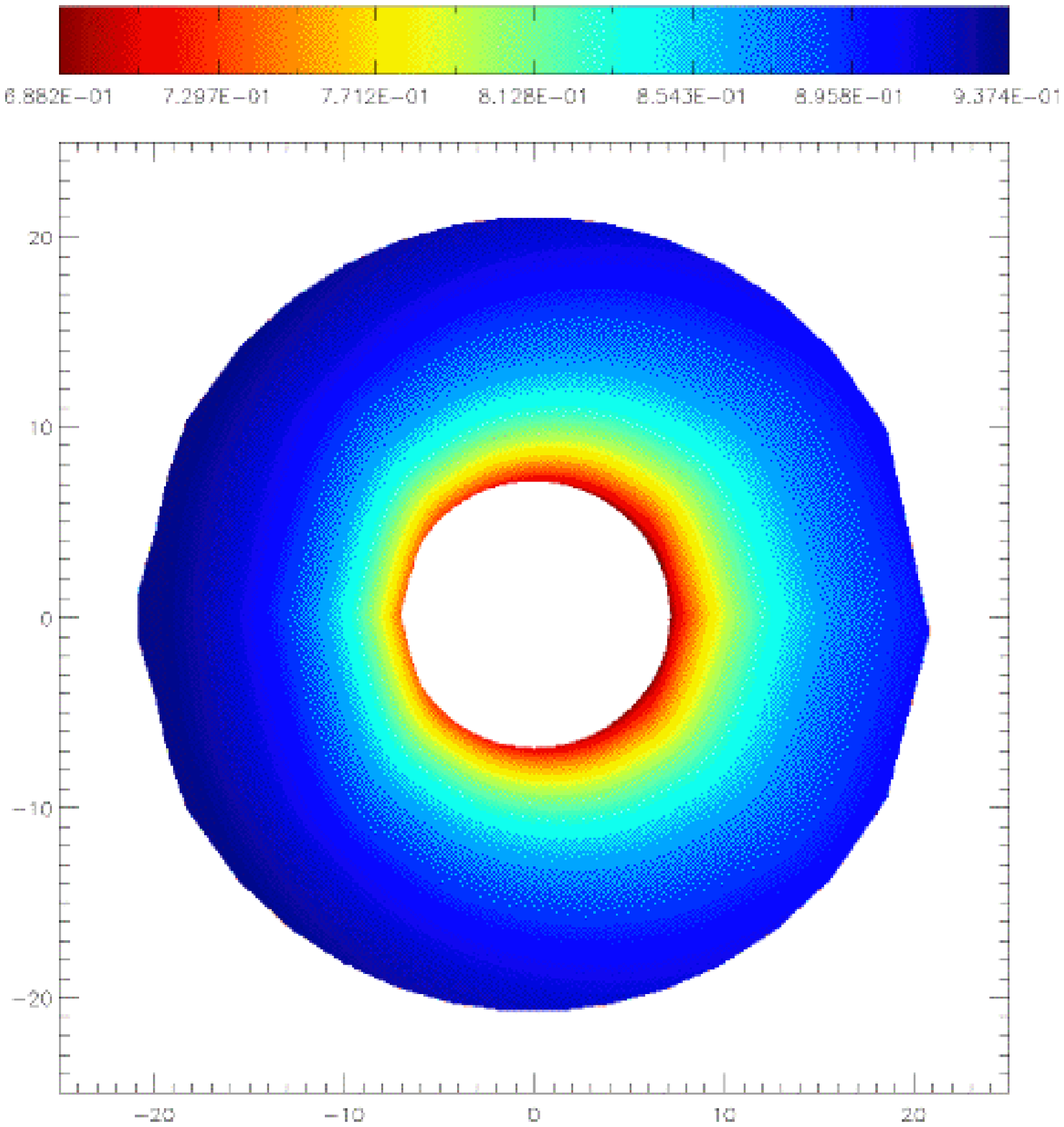}
  &
  \includegraphics[width=0.45\textwidth]{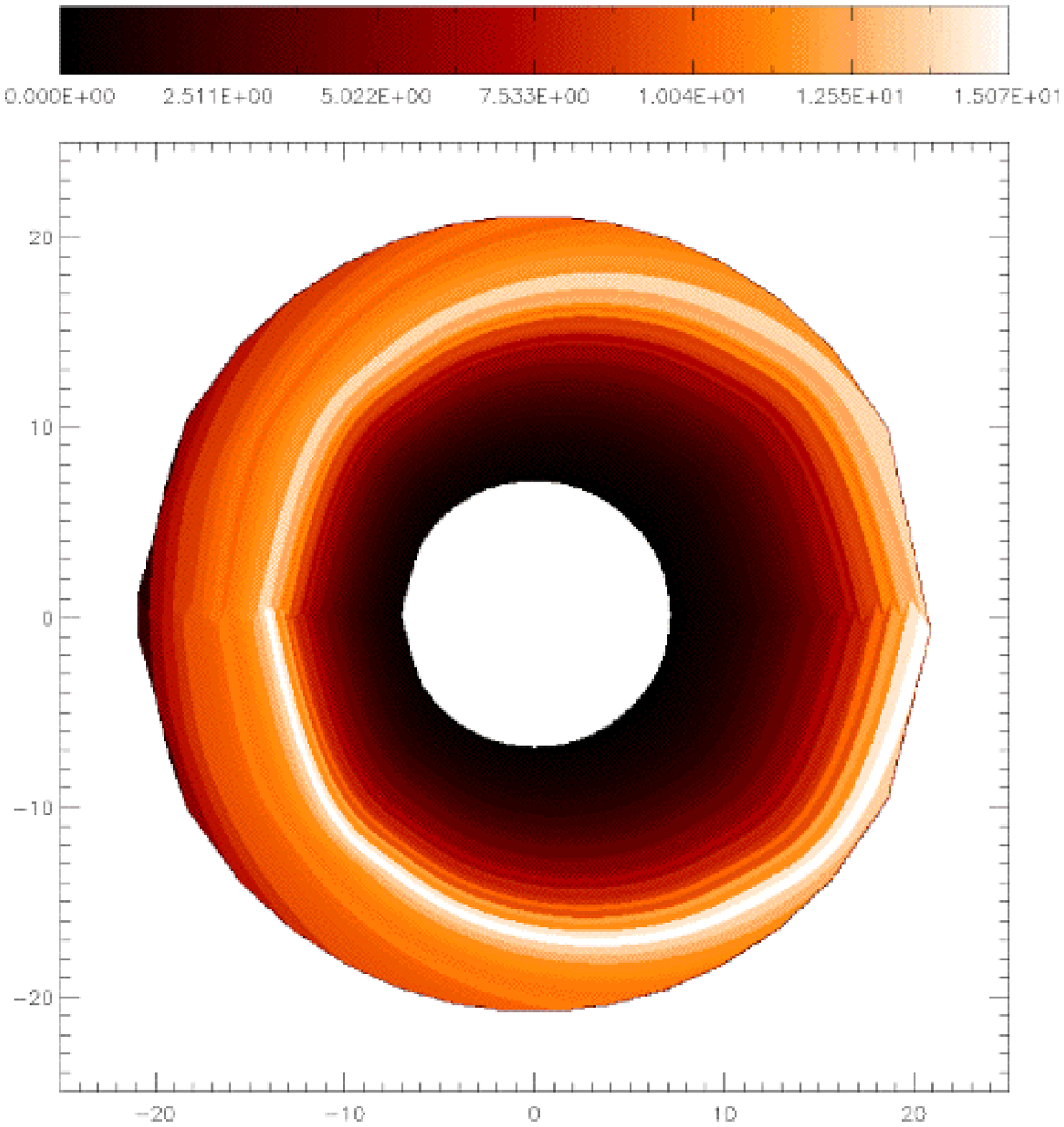}
  \\
  \includegraphics[width=0.45\textwidth]{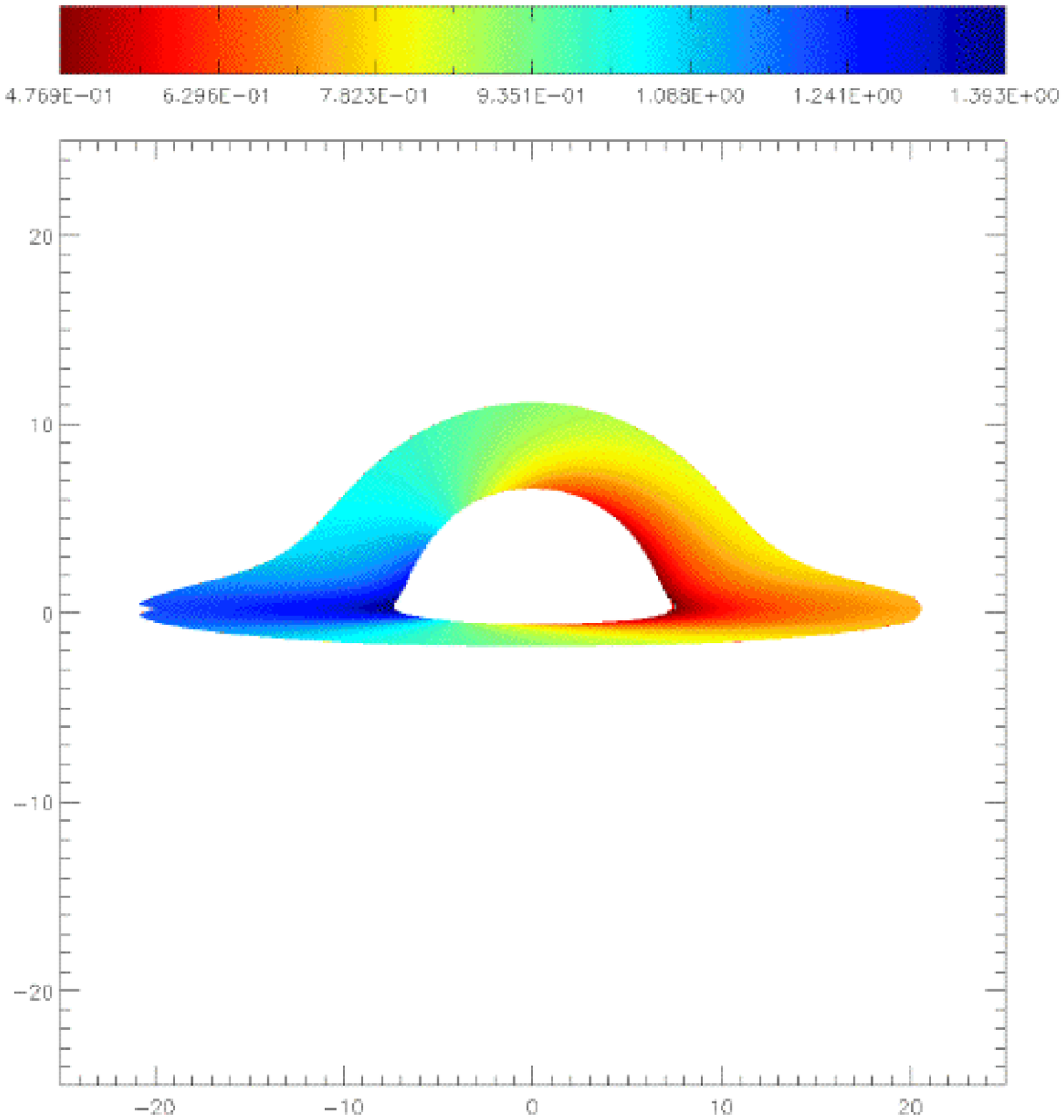}
  &
  \includegraphics[width=0.45\textwidth]{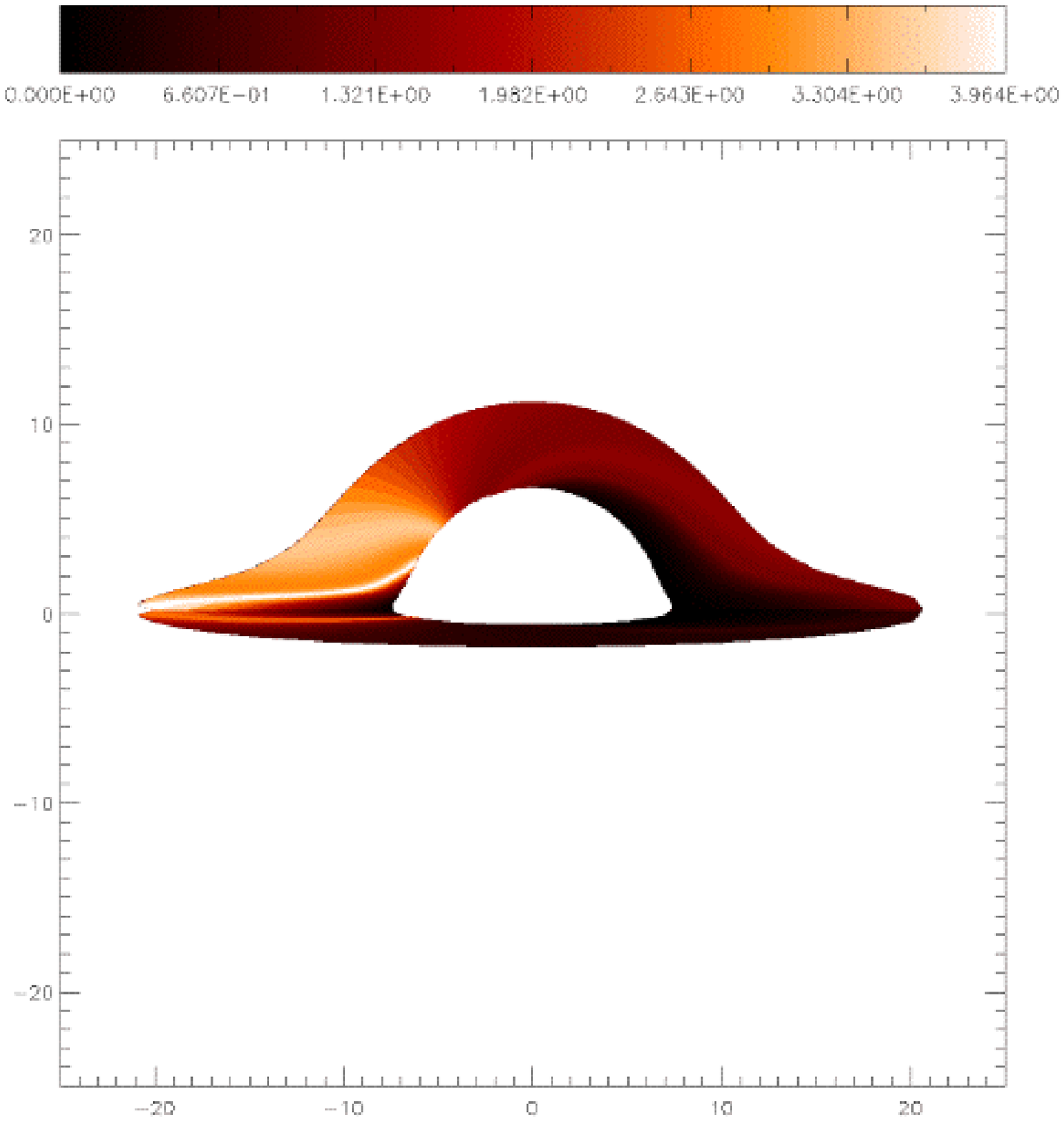}
  \end{tabular}
  \caption{Redshift images (left hand panels) and flux image (right hand panels)of 
the accretion disc on the $\left(\alpha, \beta \right)$ plane for a Schwarzschild 
black hole, for the $\theta_{o} = 30^{\circ}$ (top row) and $\theta_{o} = 
85^{\circ}$ (bottom row) cases. Redshift images are coloured by the associated 
values of $g$ as measured by the distant observer. Flux images colored by the 
$g^{4} r^{2}_{o} d\Xi = g^{4} d\alpha d\beta$ component of the relativistic line 
profile. Note the appearance of strong light bending effects in the $\theta_{o} = 
85^{\circ}$ case, as previously reported by Matt, Perola and Stella (1993), 
Zakharov and Repin (2003).}
  \label{fig:3.1.3}
  \end{center}
\end{figure*}

\begin{figure*}
  \leavevmode
  \begin{center}
  \begin{tabular}{cc}
  \includegraphics[width=0.45\textwidth]{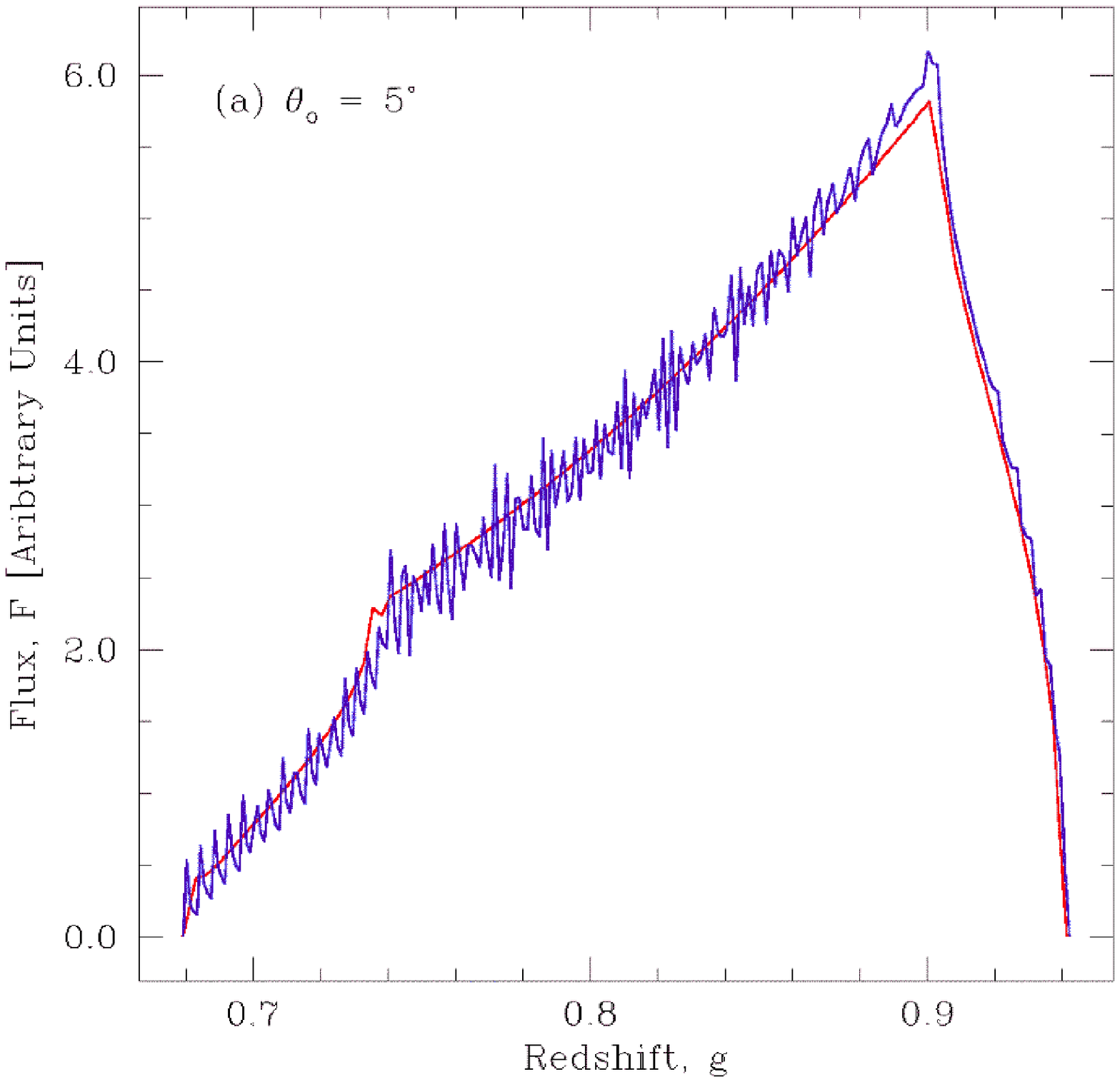}
  &
  \includegraphics[width=0.45\textwidth]{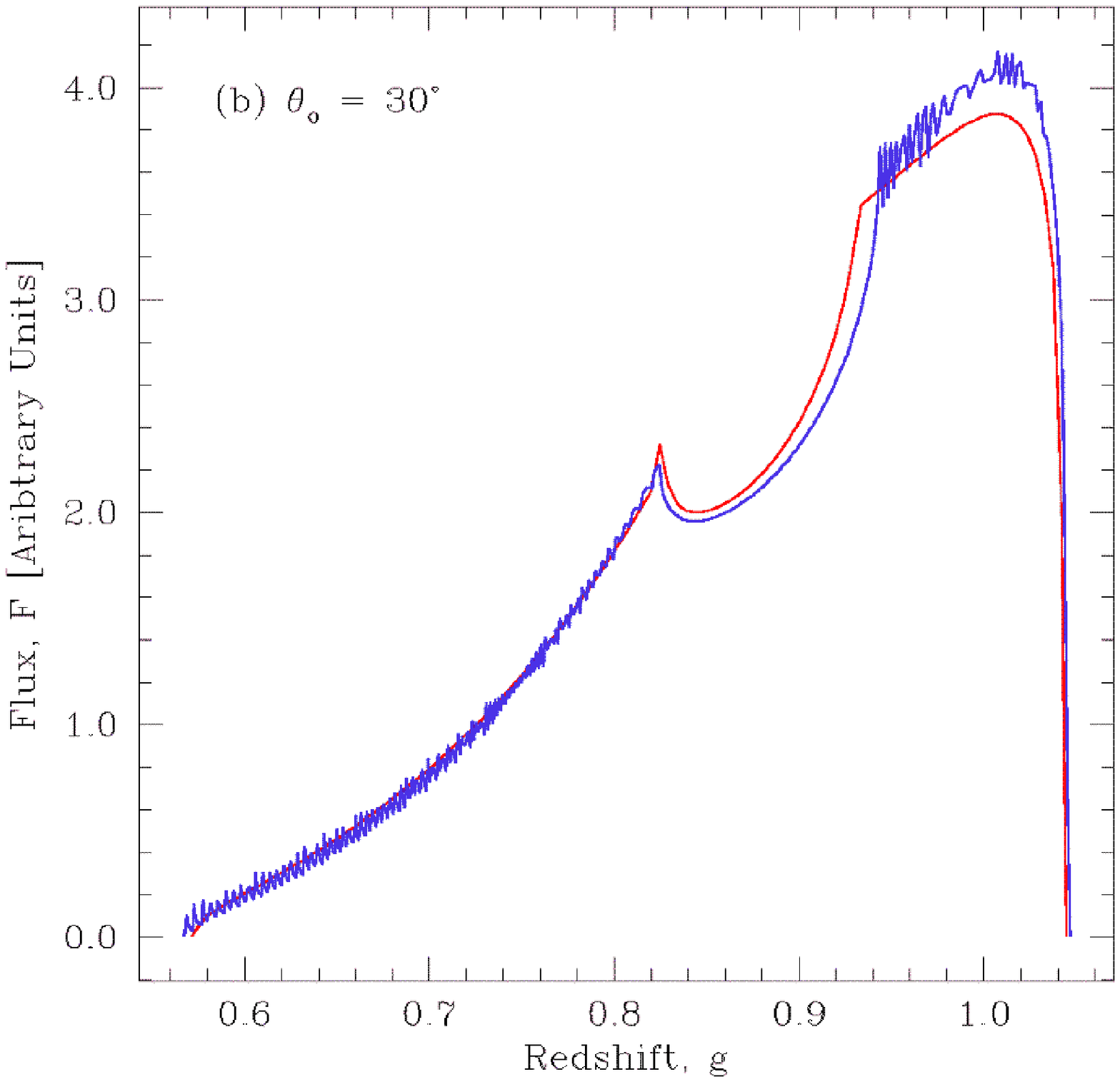}
  \\
  \includegraphics[width=0.45\textwidth]{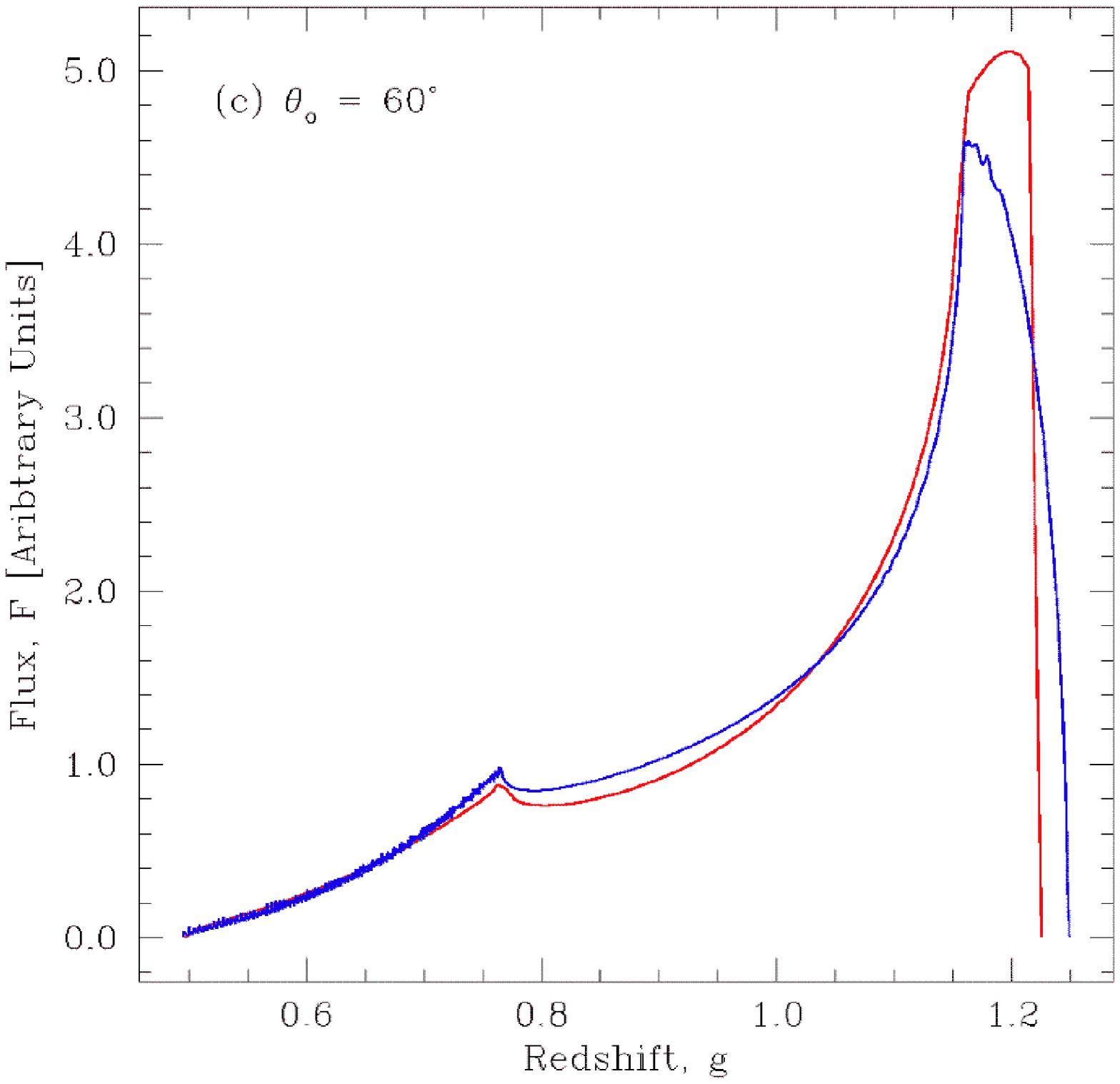}
  &
  \includegraphics[width=0.45\textwidth]{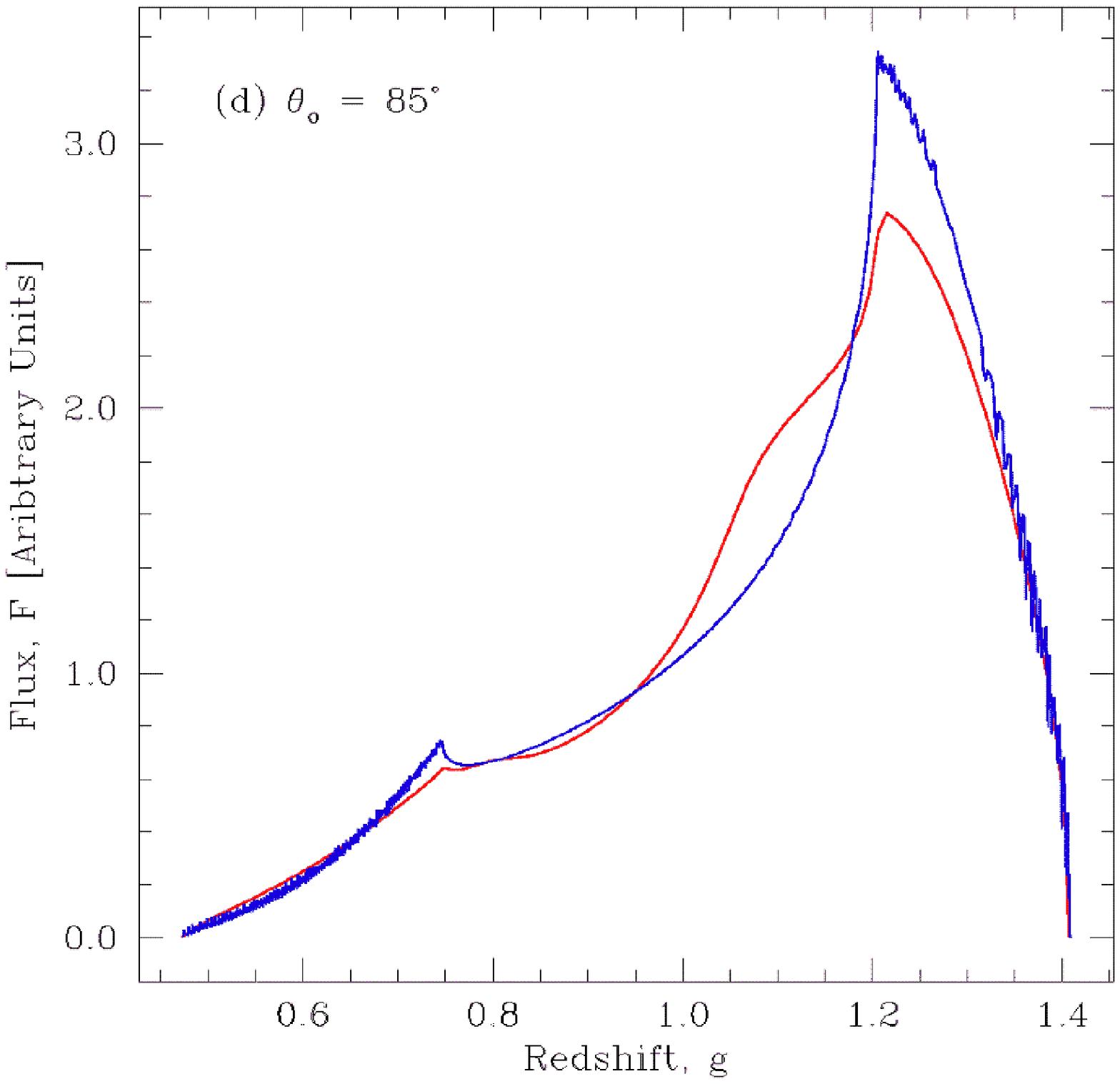}
  \end{tabular}
  \caption{Comparison of the relativistic line profile computed by our model 
({\color{red} red} solid line) with that computed by the \texttt{XSPEC diskline} 
model ({\color{blue} blue} dashed line)  for $\epsilon \left( r_{e} \right) 
\propto r^{-3}$ and $f \left( \mu_{e} \right) = 1$. At inclinations of
$< 30^{\circ}$, the profiles match to within $\sim10\%$, but the
increaing importance of lightbending (which is not included in the
{\tt diskline} code) gives a $~ 40 \%$ discrepancy in the profile shapes for inclinations 
$>60^{\circ}$. In this and all
subsequent figures the line profiles are normalised such that they
contain one photon, and all our results are unsmoothed. 
}
  \label{fig:3.2.1}
  \end{center}
\end{figure*}

\begin{figure*}
  \leavevmode
  \begin{center}
  \begin{tabular}{cc}
  \includegraphics[width=0.45\textwidth]{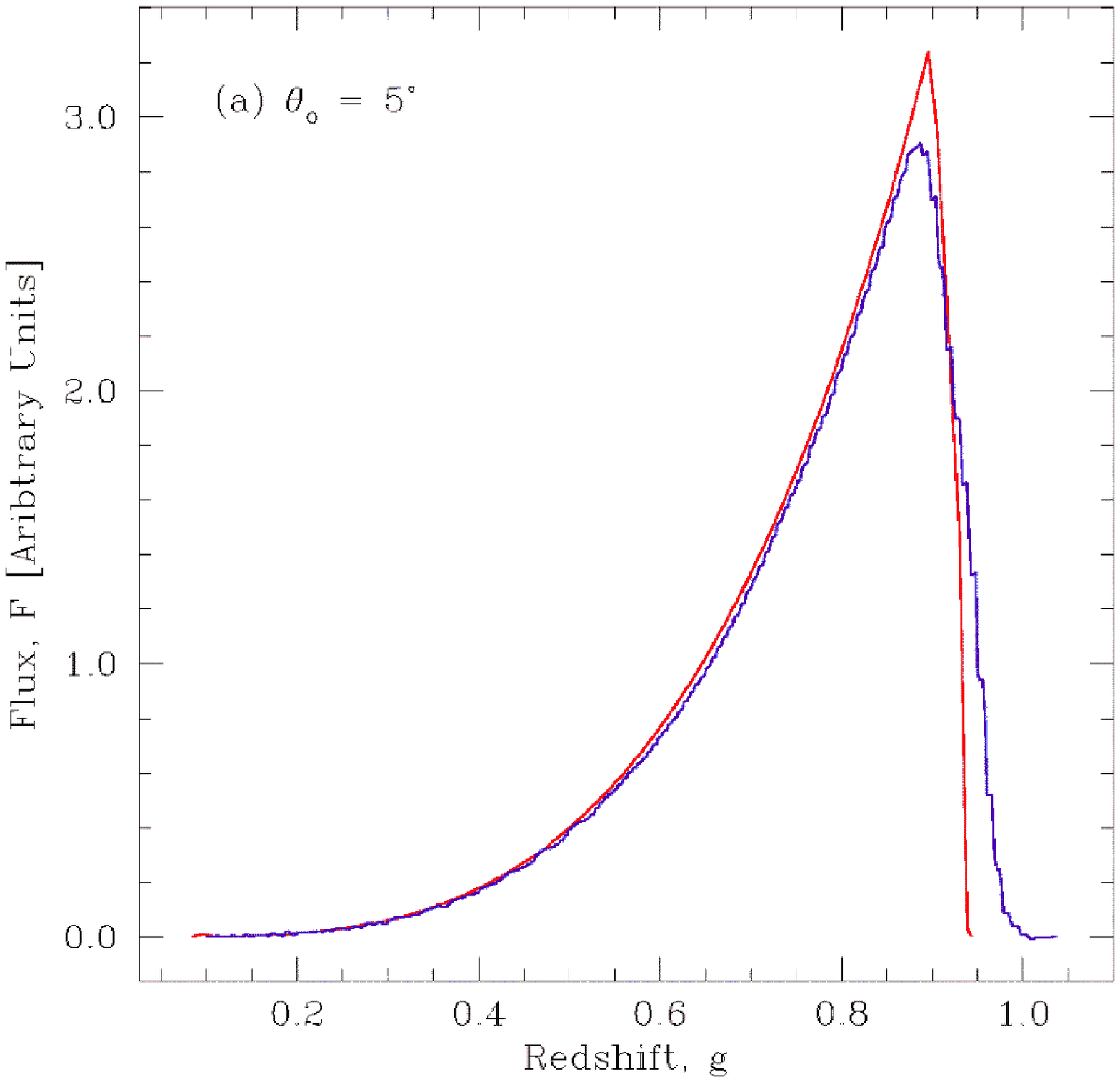}
  &
  \includegraphics[width=0.45\textwidth]{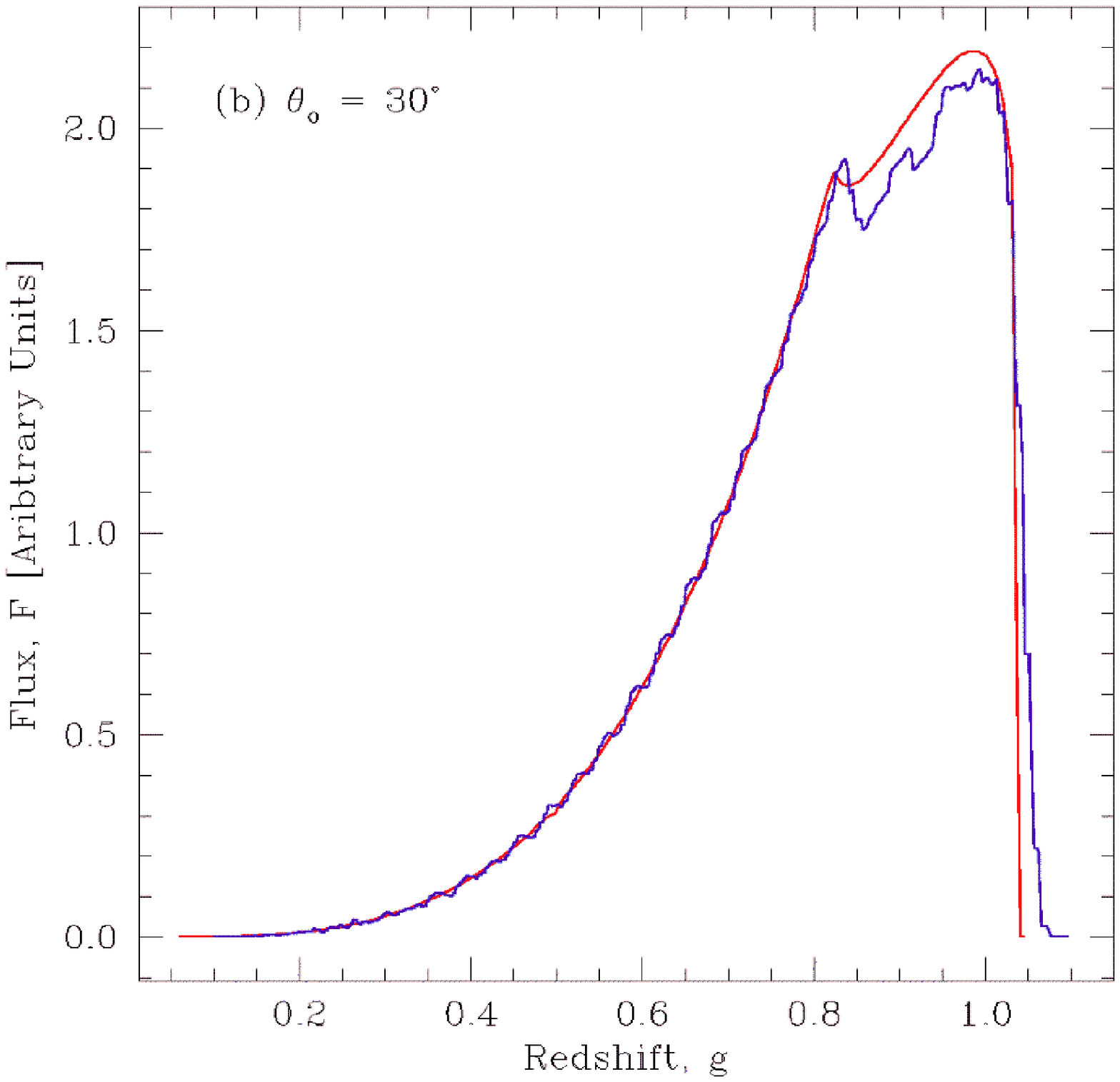}
  \\
  \includegraphics[width=0.45\textwidth]{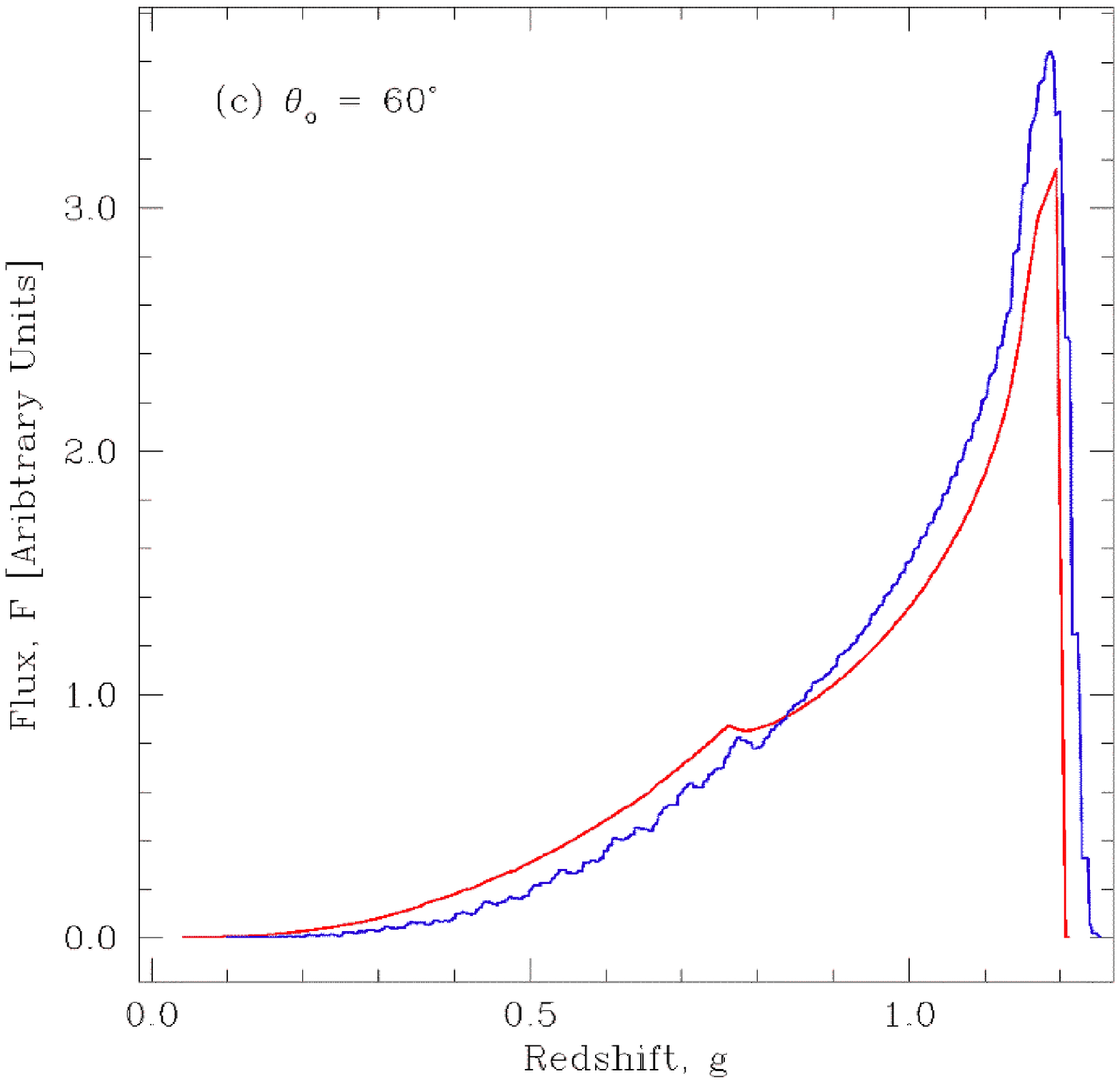}
  &
  \includegraphics[width=0.45\textwidth]{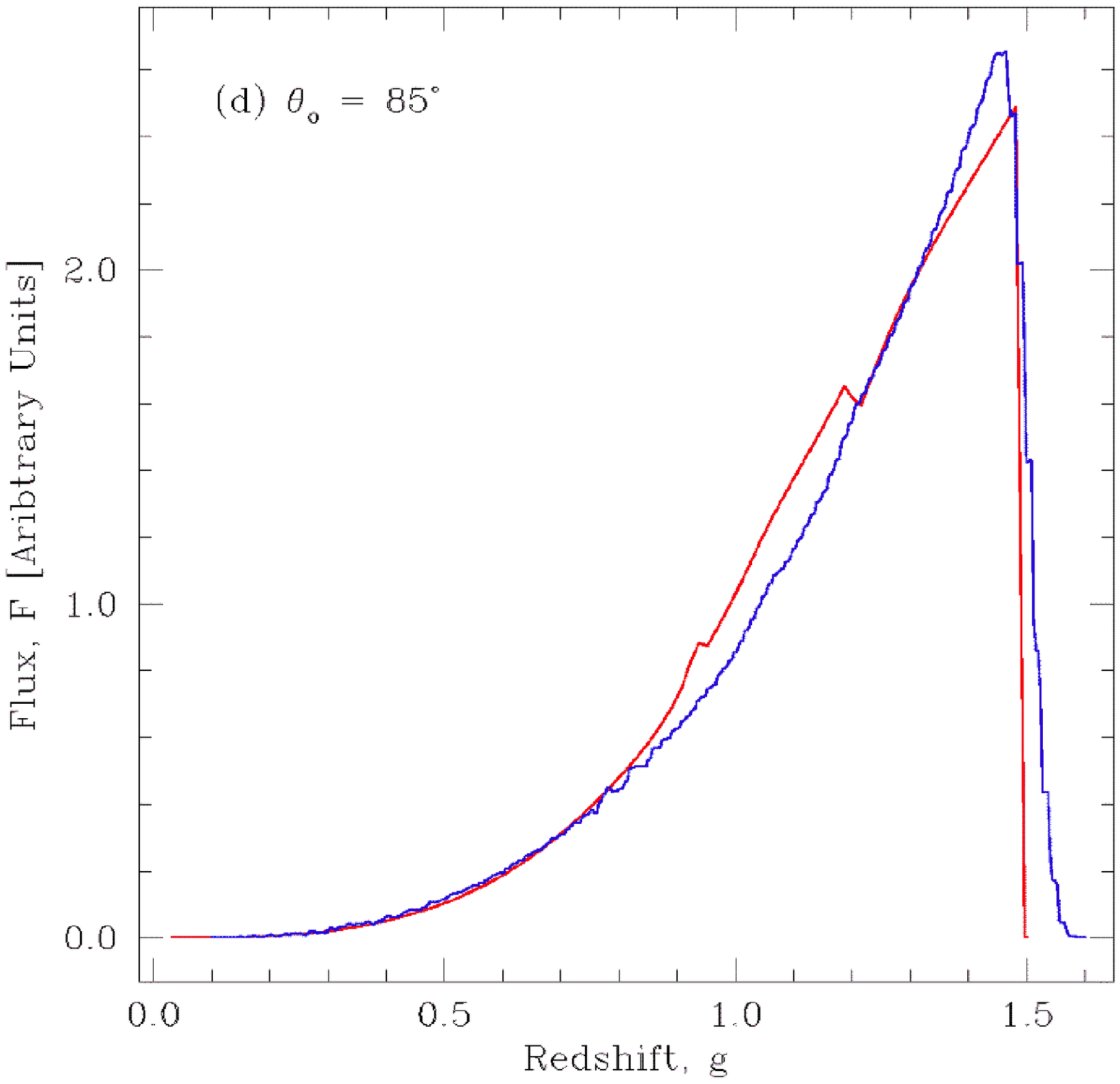}
  \end{tabular}
  \caption{Comparison of the relativistic line profile computed by our model 
({\color{red} red} solid line) with that computed by the \texttt{XSPEC laor} model 
({\color{blue} blue} dashed line)  for $\epsilon \left( r_{e} \right) \propto 
r^{-3}$ and $f \left( \mu_{e} \right) \propto \left(1+2.06\mu_e
\right)$.  The 
profiles produced by the two models match to within $~5 - 10\%$.}
  \label{fig:3.3.1}
  \end{center}
\end{figure*}

We have taken a disc from $r_{ms}$ to $r_{max} = 20
r_{g}$ (beyond which strong gravitational effects become of
diminishing importance) for both the Schwarzschild ($a = 0$, $r_{ms} = 6 r_{g}$) 
and
maximal Kerr ($a = 0.998$, $r_{ms} = 1.235 r_{g}$) cases for $\theta_{o} = 
5^{\circ} , 30^{\circ}, 
60^{\circ}$ and $85^{\circ}$. 

We first consider the extent of lightbending effects in a Schwarzschild spacetime. 
Figure \ref{fig:3.1.1} shows the three-dimensional surface in $\left( \mu_{e}, 
r_{e}, g \right)$ for
the complete set of light travel paths connecting the accretion disc to the 
observer. There is a considerable range of $\mu_{e}$ contributing to the observed 
emission at all
inclinations. For low inclinations the effect is fairly uniform, with
each radius contributing a similar range in $\mu_e$, but with a
systematic shift to larger emission angles (smaller $\mu_e$) with smaller radii. 
By contrast, at higher inclinations the lightbending is strong enough
to gravitationally lens the far side of the disc. This leads to a much
larger range of $\mu_e$ which contribute to the disc image at small radii.
In all cases, lightbending means there {\em is} a range of $\mu_e$ which
contribute to the observed disc emission, so that in 
general, the line profile {\em will} depend on the angular distribution of
the emitted flux. 

Fig. \ref{fig:3.1.2} shows the corresponding surfaces for the extreme Kerr
case. The disc now extends down to $1.235r_{g}$, far closer to the corresponding 
event horizon than in the Schwarzschild case. This introduces a greater complexity 
to the geodesic surfaces. The range of emission angles is from zero to unity in 
all cases, including the nearly face on disc at $5^{\circ}$, which has important 
consequences for the calculation of the line profile.
  
To construct the relativistic line profile, we map these surfaces on
to the $\left( \alpha, \beta \right)$ plane as discussed in the previous section,
forming images of the accretion disc, as have been previously calculated by e.g. 
\cite{CB73,L79,HM97,F97,FMA00}. In Figure \ref{fig:3.1.3} we  present  images 
of the Schwarzchild disc for the $30^{\circ}$ (top row) and $\theta_{o} = 85^{\circ}$ 
(bottom row) cases (others provide no new information qualitatively). Images on 
the left-hand side of the figure are coloured by values of the redshift factor, 
$g$, as defined by the scale at the top of each image. By contrast,  images on the 
right-hand side of the figure has each redshift bin coloured by its area on the 
observers sky, i.e. $g^{4} d\alpha d\beta$, again  with the scale defined at the 
top of each image. Strong gravitational lensing effects can now clearly be seen in 
the high inclination images. Photons from the far side of the disc pass close to 
the black hole, so the disc image is strongly distorted (\citealt{MPS93,ZR03}). 
Since the area of the disc is magnified, its contribution to the observed flux 
should be large. 
However, we stress that the low inclination images {\em are} also affected by 
lightbending (see Figure \ref{fig:3.1.1}a and \ref{fig:3.1.2}a), though they are 
not magnified by gravitational
lensing.
  
The form of the line profile is now determined from the flux image (representing
the effects of strong gravity), together with the assumed form for the
emissivity (determined by the energy release and radiative transfer
processes), which is generally taken as (ignoring azimuthal dependence):
\begin{align}
  \label{eqn:3.1.1}
    \varepsilon \left( r_{e}, \mu_{e} \right) = \epsilon \left( r_{e} \right) f 
\left( \mu_{e} \right)
\end{align}
While the flux image is a difficult numerical problem, it
depends on well known {\em physics}. By contrast, the emissivity laws
considered have rather simple forms, but are determined by the poorly
known {\em astrophysics} of the disc. Of course, there are many other outstanding 
theoretical issues that can produce a substantial impact on the line profile, 
including (but not limited to) returning radiation or lightbending that can enhance the emissivity 
of the inner part of the disc (\citealt{C75,LNP90,MKM00}), 
emission from the plunging region (\citealt{RB97}) and azimuthal
dependence of the emissivity (\citealt{CCF03,KMS01}). 
However, these are outside the scope of the current 
work.

\subsection{Comparison with the Diskline Model}

The {\tt diskline} code assumes a Schwarzchild metric ($a=0$) and additionally 
that light travels in straight lines (so the angular emissivity term is 
irrelevant). In its {\tt XSPEC} implementation it allows both arbitrary
power law $\epsilon \left( r_{e} \right) \propto r^q$ and point source
illumination. However, its analytic structure means that any radial emissivity law 
is easy to incorporate. We choose to use $q=-3$, as this is approximately the form 
of the gravitational energy release per unit disc area (see e.g. \citealt{ZDS98}).

Figure \ref{fig:3.2.1} shows our line profiles assuming $f(\mu_{e})=1$
(no angular dependance of the emissivity) compared with those from the
{\tt diskline} code. We see that our new model matches very closely to
the \texttt{XSPEC diskline} model for a nearly face on disk. Whilst
the key difference between our model and \texttt{diskline} is the
inclusion of light-bending effects, the impact of this is small at low
inclinations if there is no angular dependance to the emissivity
(but see Section 3.4).

By contrast, at high inclinations, lightbending not only means that the
line is formed from many different $\mu_{e}$, but gravitational lensing
enhances the flux from the far side of the disc.  This lensing effect
gives clear differences between our model and \texttt{diskline}. The
lensing magnifies the image of the far side of the disc, which has
velocity mostly tangential to the line of sight, so is not strongly
doppler shifted. This boosts the line profile at $g\sim 1$ (see
\citealt{MPS93}). Since the line profiles are all normalised to a
single photon, then this also makes the blue peak smaller.

In summary, the \texttt{diskline} model as incorporated into {\sc
XSPEC} produces line profiles which are accurate to $\sim 10$\% for
inclinations of less than $30^\circ$. Obviously, if the inner disc
edge $r_{min} > r_{ms}$ then the lightbending effects become
correspondingly smaller and the match between the two codes becomes
even closer.  At higher inclinations the differences between {\tt
diskline} and our code become larger due to the effects of
gravitational lensing, which leads to an effective redistribution in
flux between the blue peak and the center of the line compared to that
predicted from straight light travel paths.

\subsection{Comparison with the Laor Model}

By contrast, the \texttt{laor} code is written for extreme Kerr, and
includes a standard limb darkening law $f(\mu_e)\propto (1+2.06\mu_e)$.
The code is based on a series of photon trajectory calculations, where the disk is 
split up
into a set of rings of width $dr_{e}$ at $r_{e}$. Each part of
the ring radiates with total emissivity (radial plus angular) given
simply by the limb darkening law (i.e. no radial dependance, $q=0$)
and the line profile from that ring is built up from many light travel
paths which connect the disc to the observer. This produces a series of {\em 
transfer functions}
$T(r_{e},E_o-gE_{int})$ at each radius, analogous to Figure \ref{fig:3.1.3}a--d 
but
including the limb darkening law. These tabulated transfer functions
are read by the \texttt{laor} code in {\sc XSPEC} and used to 
build a total line profile for any given radial emissivity
$F_{o}(E_{o})=\int\varepsilon(r_{e})T(r_{e},E_{o}-gE_{int})r_{e}dr_{e}dg$.

We compare this with our code, using a $q=-3$ emissivity for both as
in the \texttt{diskline} comparisons above. We include the same limb
darkening law as used by \texttt{laor} and the results (Figure
\ref{fig:3.3.1})
show that  the overall match between our code and
\texttt{laor} is good to $\sim 5-10$\%. 

\section{The Role of the Angular Emissivity and Black Hole Spin}

\begin{figure}
  \leavevmode
  \begin{center}
  \includegraphics[width=0.9\columnwidth]{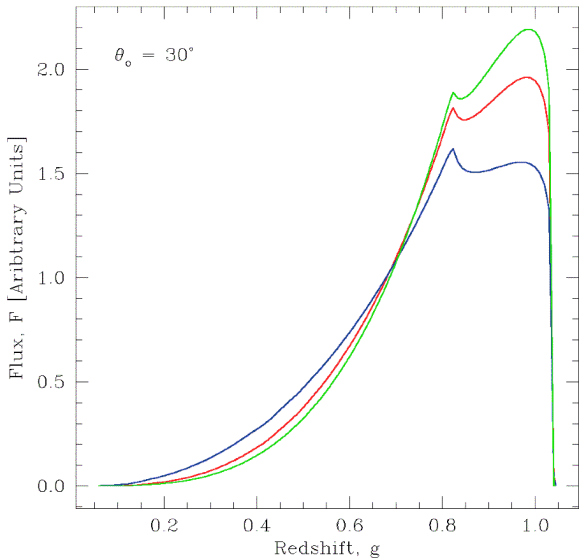}
  \caption{Comparison of the relativistic line profiles generated by our model 
with (a) $\epsilon \left( r_{e} \right) \propto r_{e}^{-3}$, $f \left( \mu_{e} 
\right) = 1$ ({\color{red} red} lines), (b) $\epsilon \left( r_{e} \right) 
\propto r_{e}^{-3}$, $f \left( \mu_{e} \right)  \propto \mu^{-1}_{e}$ 
({\color{blue} blue} lines), (c) $\epsilon \left( r_{e} \right) \propto 
r_{e}^{-3}$, $f \left( \mu_{e} \right)  \propto \left( 1 + 2.06 \mu_{e} \right)$ 
({\color{green} green} lines) for a maximal Kerr black hole with the
disc extending from $1.235 - 20 r_{g}$. The relative height of the
blue wing changes by $\sim 35$\% for different angular
emissivity laws, anti-correlated with the slope of the red wing. 
}
  \label{fig:3.4.1}
  \end{center}
\end{figure}

\begin{figure}
  \leavevmode
  \begin{center}
  \includegraphics[width=0.9\columnwidth]{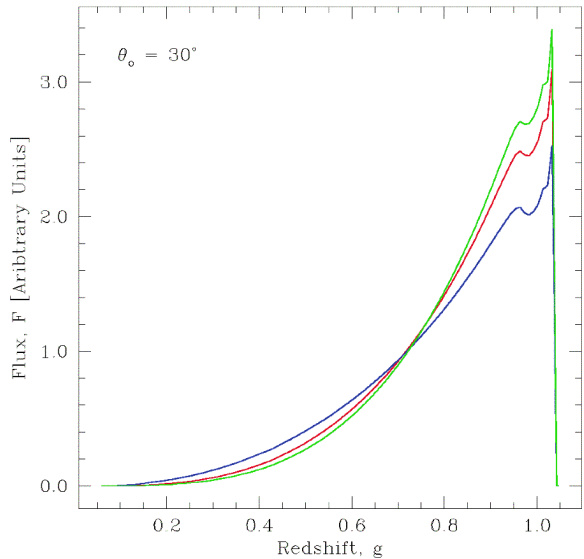}
  \caption{As in Fig. \ref{fig:3.4.1} with the disc now extending from
$1.235 - 400 r_{g}$. There is still a $\sim$ 25\% change in the blue
wing height and significant change in red wing slope for the different
angular emissivities, despite the inclusion of the outer disc
regions.
}
  \label{fig:3.4.2}
  \end{center}
\end{figure}

\begin{figure}
  \leavevmode
  \begin{center}
  \includegraphics[width=0.9\columnwidth]{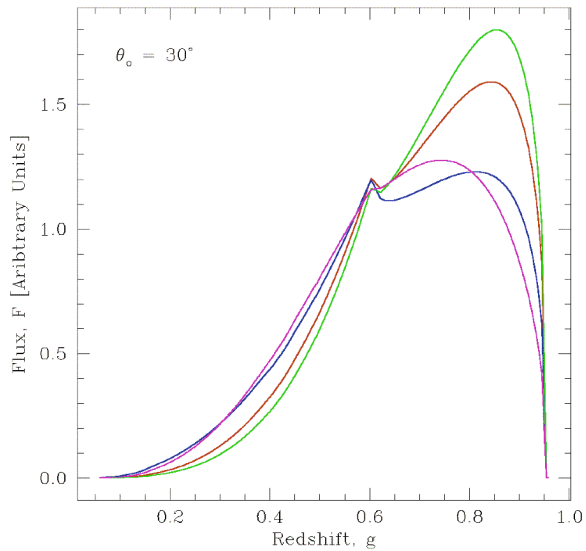}
  \caption{As in Fig. \ref{fig:3.4.1} with the disc now extending from
$1.235 - 6 r_{g}$. The additional {\color{magenta} magenta} line
is for a limb darkened angular emissivity with more centrally
concentrated radial emissivity, $\propto r_{e}^{-4.5}$. This is very
similar to the  {\color{blue} blue} line profile derived from a very
different radial emissivity, $\propto r_{e}^{-3}$, with a limb
brightened angular emissivity.
}
  \label{fig:3.4.3}
  \end{center}
\end{figure}

\begin{figure}
  \leavevmode
  \begin{center}
  \includegraphics[width=0.9\columnwidth]{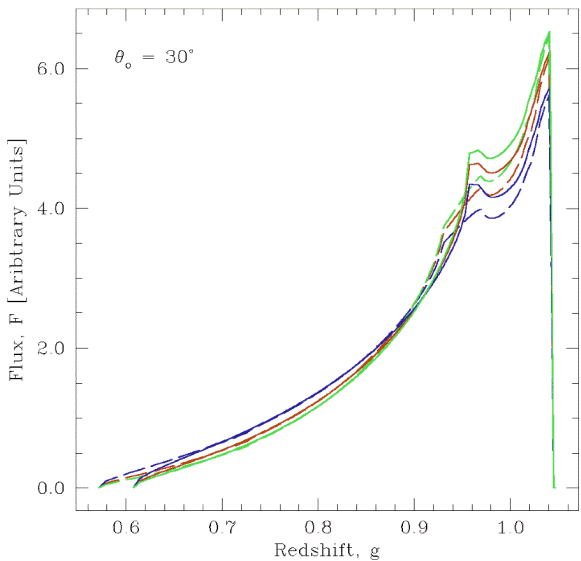}
  \caption{As in Fig. \ref{fig:3.4.1} but with the disc extending from
$6-400 r_g$ in an extreme Kerr (solid line) and Schwarzchild (dashed
line) spacetime.  The differences between the line profiles produced
for the same sized disc in different assumed spacetimes is of order $\sim 5$\% for 
a given angular emissivity. The effect of changing the angular
emissivity is also similarly small ($\sim 5-10$\%). This
contrasts with the much larger effects seen in the extreme Kerr metric
for a  disc extending down to 1.235$r_g$, where lightbending is much
more important (Fig. \ref{fig:3.4.1}).
}
  \label{fig:3.4.4}
  \end{center}
\end{figure}

\begin{figure}
  \leavevmode
  \begin{center}
  \includegraphics[width=0.9\columnwidth]{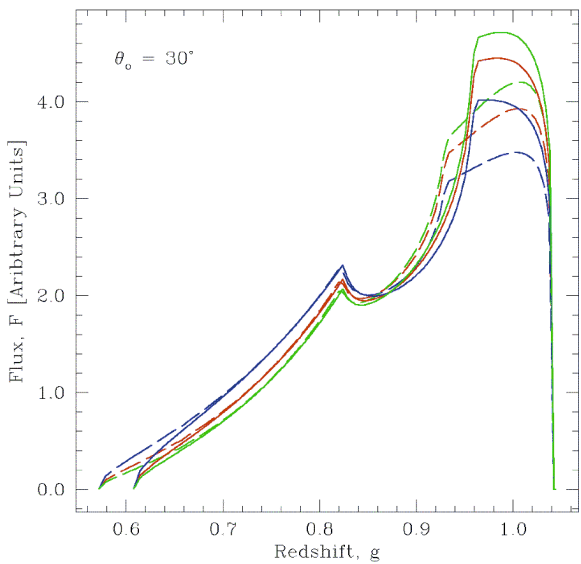}
  \caption{As in Fig. \ref{fig:3.4.4} but with the disc extending from $6 - 20 r_{g}$.  
The differences now are of order 15-20\%.
}
  \label{fig:3.4.5}
  \end{center}
\end{figure}

The effect of appling a radial emissivity is straightforward. The
transfer function describing all the relativistic effects from a given
radial ring of the disc is unaffected, so the effect 
is simply to change the weighting of the line profile from
each radial ring of the disc.

By contrast, the effect of the angular distribution is far more
subtle. A given radial ring on the disc can contribute to the line
profile from a range of emission angles. The relative weighting of
these is determined by the angular emissivity, so it forms part of the
calculation of the transfer function itself. 

Different angular emissivity laws can have striking effects on the
form of the relativistic line profile, which we illustrate in
Fig. \ref{fig:3.4.1} for a maximal Kerr geometry (a=0.998) with the
disc extending (as previously) from $1.235 - 20 r_{g}$ and $\theta_{o}
= 30^\circ$. The line profiles here all implement the standard radial
emissivity law of $r^{-3}$. However, we now compare a range of angular
emissivity laws, these being (from top to bottom at the blue peak in
Fig. \ref{fig:3.4.1}) the standard limb darkening law (as discussed in
\S 3.3), followed by the constant angular emissivity case (as used in
\S 3.2). An ionized disc could also be limb {\em brightened}, with the
probable limiting case of $f(\mu)\propto 1/\mu$ as expected from
optically thin material, shown as the bottom line in Fig.
\ref{fig:3.4.1}. There is a $\sim 35 \%$ difference in the height of
the blue peak depending of the form of the angular emissivity used.

However, such a limited range of radii is probably not very realistic.
The disc should extend out to much greater distances from the black
hole, where the relativistic effects (including lightbending) are less
extreme. However, realistic emissivities strongly weight the
contribution from the innermost regions, so the effective dilution of
the relativistic effects by including the outer disc is not
overwhelming. Fig. \ref{fig:3.4.2} shows the line profiles generated using
the same angular emissivity laws for a disc extending from $1.235 -
400 r_{g}$, again with $\theta_{o} = 30^\circ$. There are still 
significant differences in the line profiles, with a $\sim
25 \%$ difference in the height of the blue peak while the
red wing slope changes from $F_o(E_o) \propto E_o^{3.5}$ (limb darkened) to 
$\propto E_o^{2.5}$ (limb brightened). 

Despite the expectation of an extended disc, some recent
observational studies (e.g. \citealt{R04}) have tentatively suggested
that the disc is very small, from $\sim 1.235-6 r_{g}$. This enhances
the importance of lightbending. Fig. \ref{fig:3.4.3} shows the line
profiles for a disc extending from 1.235--6$r_{g}$ using the different
angular emissivity laws of Fig.
\ref{fig:3.4.1}. The blue peak height differences are $\sim$
40\%, and the red wing slopes are different.
For comparison we also show a limb darkened profile obtained
from a very different {\em radial} emissivity of $r^{-4.5}$. This is
very similar to the extreme limb brightened profile obtained from the
$r^{-3}$ radial weighting.  We caution that uncertainties in the
angular distribution of the line emissivity can change the expected
line profile due to lightbending effects even at low/moderate
inclinations, and that this can affect the derived radial emissivity.

Currently, the only available models in {\tt XSPEC} have either zero
or maximal spin. A zeroth order approximation to spacetimes with
different spins is to use the maximal Kerr results but with a disc
with inner radius given by the minimum stable orbit for the required
value of $a$ (e.g. \citealt{L91}). We test this for the most extreme
case of $a=0$ modelled by a maximal Kerr spacetime with
$r_{min}=6r_g$. Fig. \ref{fig:3.4.4} compares this with a true
Schwarzschild calculation for a disc extending from $6 - 400 r_{g}$
with $\theta_{o} = 30^{\circ}$ for a range of angular emissivities.
The differences between the spacetimes (for a given angular
emissivity) are at most $\sim 5\%$. This is roughly on the same order as
the effect of changing the angular emissivity, which is much reduced
here compared to Fig. \ref{fig:3.4.1} due to the larger $r_{min}$.
Assumptions about both spin and angular emissivity become 
somewhat more important for smaller outer disc radii. 
Fig. \ref{fig:3.4.5} shows this for a disc between $6 - 20 r_{g}$
(directly comparable to Fig. \ref{fig:3.2.1}).

\section{Conclusions}

Recent observational studies have provided evidence for highly
broadened fluorescent iron K$\alpha$ lines. While there are a variety
of line profiles seen (e.g. \citealt{LZ01}), there are some objects
where the line implies that there is material down to the last stable
orbit in a maximally spinning Kerr spacetime (most notably
MCG-6-30-15: \citealt{W01}). However, the strong gravity codes
generally used to model these effects are now over a decade
old. Increased computer power means that it is now possible to improve
on these models. We describe our new code to calculate these effects,
which uses uses fully adaptive gridding to map the image of the disc
at the observer using the analytic solutions of the light travel
paths. This is a very general approach, so the code can easily be
modified to incorporate different emission geometries.

We compare the results of our new code with those from {\tt diskline}
and {\tt laor} (publically available in the {\sc XSPEC} spectral
fitting package) for Schwarzchild and extreme Kerr spacetimes.  These
previous models are accurate to $\sim 10\%$ with realistic ($\propto
r^{-3}$) radial emissivities.  However, they make specific assumptions
regarding the angular dependence of the emitted flux, which may or may
not be valid. Lightbending is {\em always} important for a disc which
extends down below 20$r_g$, in that the image of the disc at the
observer {\em always} consists of a range of different emission
angles. This can produce significant changes to the derived line
profile, especially in extreme Kerr spacetimes. Whilst calculating
strong gravitational effects is a difficult numerical problem, the
underlying physics is well known. By contrast, the {\em angular}
emissivity is an astrophysical problem, and is not at all well known
as it depends on the ionization state of the disc as a function both
of height and radius. Before we can use the line profiles to provide a
sensitive test General Relativity and probe the underlying physics, we
will need to have a much better understanding of the astrophysics of
accretion.

This code will be publically released for inclusion as a convolution
model in the \texttt{XSPEC} spectral fitting package. This will include
arbitrary spin and  inner and  outer disc radii as well as allowing both angular
and radial emissivities to be specified.

After this paper was submitted we learnt of the independent work by 
\citep{DKY04} which also develops a new strong gravity code. Their
results match very closely with those presented here.

\section*{Acknowledgements}

We are grateful to an anonymous referree for useful comments and
suggestions on previous versions of this manuscript.
We would also like to thank E. Agol and M. Calvani for
useful discussions and encouragement.

%%%%%%%%%%%%%%%%%%%%%%%%%%%%%%%%%%%%

% \bsp % ``This paper has been produced using the ...''

\label{lastpage}

\end{document}